%% file: paper.tex
\documentclass{aa}  

\usepackage{graphicx}
\usepackage{txfonts}
\graphicspath{{./}{figures/}}

\begin{document} 

\title
{Study of the 2024 major Vela glitch at the Argentine Institute of Radioastronomy}




\author{
Ezequiel Zubieta\inst{1}\and
Ryan Missel\inst{2}\and
Susana B. Araujo Furlan\inst{3,4}\and
Carlos O. Lousto\inst{5,6}\thanks{E-mail: colsma@rit.edu (COL)}\and
Federico Garc\'{i}a\inst{1,7}\and
Santiago del Palacio\inst{1,8}\and
Guillermo Gancio\inst{1}\and
Jorge A. Combi\inst{1,7}\and
Linwei Wang\inst{2}}

\institute{Instituto Argentino de Radioastronom\'ia (CCT La Plata, CONICET; CICPBA; UNLP), C.C.5, (1894) Villa Elisa, Buenos Aires, Argentina.\and
Golisano College of Computing and Information Sciences, Rochester Institute of Technology Rochester, NY 14623, USA\and
Facultad de Matemática, Astronomía, Física y Computación, UNC. Av. Medina Allende s/n , Ciudad Universitaria, CP:X5000HUA - Córdoba, Argentina.\and
Instituto de Astronom\'\i{}a Teórica y Experimental, CONICET-UNC, Laprida 854, X5000BGR – Córdoba, Argentina \and
Center for Computational Relativity and Gravitation, Rochester Institute of Technology, 85 Lomb Memorial Drive, Rochester,New York 14623, USA\and
School of Mathematical Sciences, Sciences Rochester Institute of Technology Rochester, NY 14623, USA\and
Facultad de Ciencias Astron\'omicas y Geof\'{\i}sicas, Universidad Nacional de La Plata, Paseo del Bosque, B1900FWA La Plata, Argentina\and
Department of Space, Earth and Environment, Chalmers University of Technology, SE-412 96 Gothenburg, Sweden}




\abstract
{We report here on new results of the systematic monitoring of southern glitching pulsars at the Argentine Institute of Radioastronomy. In particular, we study in this work the new major glitch in the Vela pulsar (PSR J0835$-$4510) that occurred on 2024 April 29.}
{We aim to thoroughly characterise the rotational behaviour of the Vela pulsar around its last major glitch and investigate the statistical properties of its individual pulses around the glitch.}
{We characterise the rotational behaviour of the pulsar around the glitch through the pulsar timing technique. We measured the glitch parameters by fitting timing residuals to the data collected during the days surrounding the event. In addition, we study Vela individual pulses during the days of observation just before and after the glitch. We selected nine days of observations around the major glitch on 2024 April 29 and studied their statistical properties with the Self-Organizing Maps (SOM) technique. We used Variational AutoEncoder (VAE) reconstruction of the pulses to separate them clearly from the noise.}
{We obtain a precise timing solution for the glitch. We find two recovery terms of $\sim 3~\mathrm{days}$ and $\sim 17~\mathrm{days}$. We find a correlation of high amplitude with narrower pulses while not finding notable qualitative systematic changes before and after the glitch.}
{}

\maketitle


\section{Introduction}\label{sec:intro}

Pulsars are highly magnetised neutron stars emitting beams of electromagnetic radiation from their magnetic poles. As these beams sweep across our line of sight, we observe regular pulses of emission, with a frequency corresponding to the star's rotational frequency. Pulsars are extremely dense compact objects, which provides them with a very large moment of inertia. Their rotation is extraordinarily stable, which makes them in some cases as accurate as atomic clocks \citep{2012MNRAS.427.2780H}. However, some young pulsars present abrupt changes in their rotational frequency known as glitches. Currently, close to 200 pulsars are known to glitch \citep{Espinoza:2011pq,Yu2013,Manchester:2018jhy}.

The Vela pulsar (PSR J0835--4510) was the first pulsar known to glitch \citep{1969Natur.222..228R,1969Natur.222..229R}. Nowadays, 26 glitches have been reported in the Vela pulsar \citep{2022MNRAS.510.4049B, 2024ATel16608....1Z}. Glitches are mainly characterised by the relative increase in the frequency of the pulsar ($\Delta\nu/\nu$). Giant glitches have a typical size of $\Delta\nu/\nu \sim10^{-6}$ while small glitches have a typical size of $\Delta\nu/\nu \sim10^{-9}$. The Vela pulsar is one the most studied pulsars, given that it is the brightest pulsar from the southern hemisphere and presents giant glitches quasi-periodically, with these giant glitches occurring every 2 to 3 years. 

Although the dynamics of the glitches and the mechanisms that may trigger them are poorly understood, it is widely accepted that they are a consequence of the interaction between the superfluid interior of neutron stars and their solid crusts \citep{2012PhRvL.109x1103A,2013PhRvL.110a1101C,2015IJMPD..2430008H}. Therefore, observations of glitches are crucial for probing the internal structure of neutron stars and provide information on their equation of state \citep{1992RSPTA.341...29L, 2017JPhCS.932a2037G}. The glitch magnitude \citep{1999PhRvL..83.3362L} can be used to estimate the moment of inertia of the superfluid component of the star and also to estimate the total mass of the neutron stars \citep{2015SciA....1E0578H,2020MNRAS.492.4837M,khomenko_haskell_2018}. In addition, post-glitch relaxation can provide information on the mutual friction between the vortices in the superfluid and the solid crust \citep{2018ApJ...865...23G}. Finally, changes in the pulsar emission previous or during the glitch can provide further information of the coupling between the dynamics of the neutron stars and the dynamics of their magnetospheres \citep{2020ApJ...897..173B}. In particular, during the 2016 Vela glitch, \cite{2018Natur.556..219P} found that the pulse shape became broader, and that a null state occurred just prior to the glitch, accompanied by a missed pulse. This was followed by a temporary loss of linear polarisation in the subsequent pulses, lasting for a short duration around the glitch event. This was later interpreted by \cite{2017JPhCS.932a2037G} and \cite{2020ApJ...897..173B} as a quake occurring deep inside the crust that induced high-frequency oscillations leading to the observed changes in the magnetosphere. To better understand the glitch phenomenon, more real-time observations of glitches are needed. However, these events are difficult to capture in real-time due to their unpredictable occurrences. It is also a valid question whether there are any magnetospheric signals that could serve as precursors to a glitch, providing clues before the event occurs.

The Pulsar Monitoring in Argentina\footnote{\url{https://puma.iar.unlp.edu.ar}} (PuMA) collaboration has been using the two antennas from the Argentine Institute of Radio Astronomy (IAR) since 2019 to observe, with high cadence, a set of pulsars from the southern hemisphere that have exhibited glitches \citep{Gancio2020}. Given that the close follow-up of the Vela pulsar is a major goal of the collaboration, we already detected its last three giant glitches. We detected the 2019 glitch with observations three days before and three days after the epoch of the glitch \citep{atel_vela}, we first reported the 2021 glitch \citep{2021ATel14806....1S} with observations performed only one hour after the glitch, and we also first reported the 2024 giant glitch in \cite{2024ATel16608....1Z}. In addition, in \cite{2024A&A...689A.191Z} we reported a small glitch that occurred $\sim 70~\mathrm{d}$ after the 2021 giant glitch. We will continue monitoring closely the Vela pulsar with 3.66 hours (3:40~hours:minutes) daily observations in an effort to catch a glitch "live".

Given that the Vela pulsar is exceptionally bright, it is possible to study its individual pulses. In \cite{Lousto:2021dia} we analysed nine of our daily observations, each lasting over three hours and capturing approximately 120,000 pulses. We then applied machine learning techniques to investigate their statistical properties. Firstly, we utilised the DBSCAN clustering method, grouping pulses primarily by amplitude. We thus found a correlation between higher amplitudes for pulses that arrived earlier, and a weaker (polarisation-dependent) correlation with the pulse width. We also identified clusters of "mini-giant" pulses with amplitudes about ten times the average.

In a parallel analysis, we used the Variational AutoEncoder (VAE) method to reconstruct the pulse shapes \citep{kingma2014autoencoding}, effectively distinguishing them from noise. We chose one observation to train the VAE and we applied it to data from the other observations. We then employed Self-Organizing Maps (SOM) clustering techniques \citep{teuvo1988som} on these reconstructed pulses to determine four distinct clusters per day, per radio telescope. We found that the results were robust and consistent, supporting models of emission regions at different altitudes within the pulsar's magnetosphere, separated by around $100~\mathrm{km}$.

Given the success of this methodology, we applied these techniques to data surrounding the major glitch on July 22nd, 2021, with daily observations collected around the event. Our findings, reported in \cite{Zubieta:2022umm}, did not reveal any unusual behaviour based on SOM clustering analysis. In this work, we employ this same technique, applying it to a consistent set of observations from a few days before and after the 2024 glitch, in search of any distinctive emission features associated with the glitch event.

\section{Pulsars Glitch Monitoring Program at IAR}\label{sec:Pugliese}



The Argentine Institute of Radio Astronomy (IAR), which is located at latitude $-34\degr 51\arcmin 57\arcsec.35$ and longitude $58\degr 08\arcmin 25\arcsec.04$, counts with two 30~m single-dish antennas that are aligned on a North--South direction and separated by $120$~m, covering a declination range of $-90\degr < \delta <-10\degr$ and an hour angle range of two hours east/west, $-2\,\mathrm{h} <t< 2\, \mathrm{h}$.

The data are obtained with a timing resolution of $146~\mu s$ for both antennas with ETTUS receivers. For A1, we utilize 128 channels of 0.875 MHz in single (circular) polarisation mode centred at 1400 MHz, whereas for A2, we use 64 channels of 1 MHz in dual polarisation (both circular polarisations added) centred at 1428 MHz. To limit systematic effects, we observe each target separately using both antennas whenever possible. In \cite{Gancio2020} we provided a thorough explanation of the features of the front ends in A1 and A2. In addition, the analysis of the radio frequency interference (RFI) environment provided in \cite{Gancio2020} revealed that the radio band from 1 GHz to 2 GHz has a low level of RFI activity that is adequate for radio astronomy, despite the fact that the IAR is not located in a quiet zone for RFI.

In addition to the ETTUS boards, we added a parallel digitalizer board in the middle of 2022 made up of Reconfigurable Open Architecture Computing Hardware (ROACH) boards \citep{2024RMxAC..56..131G,2023BAAA...64..304A}. The ROACH boards are configured to observe, in both antennas, with dual circular polarisation and $400~\mathrm{MHz}$ of bandwidth. In this work we use the data from the ETTUS receivers given that we did not fully characterize yet the results of the timing with the ROACH receivers.

We emphasise that the major advantage of the IAR's observatory is the availability for long-term, high-cadence monitoring of bright sources. Therefore, we are carrying out since 2019 an intensive monitoring campaign of known bright pulsars in the southern hemisphere in the L-band (1400 MHz) using the two antennas. Our observational schedule is focused on high-cadence observations with up to daily cadence for some pulsars and reaching observations as long as 3.66 hours per day.
This observational program clearly results in an unique database built to explode the benefits of high-cadence monitoring, and increases the probability of observing a glitch "live," an objective that has only rarely been met by other monitoring programs \citep[e.g.][]{2018Natur.556..219P,Flanagan1990RapidRO,2002ApJ...564L..85D}.

The findings of our campaign include: (i) the report of the 2021 and 2024 glitches in the Vela pulsar \citep{2021ATel14806....1S, 2024RMxAC..56..161Z, 2024ATel16608....1Z}, alongside the confirmation of its 2019 glitch \citep{2019ATel12482....1L} and the discovery of a minor glitch near the 2021 giant glitch \citep{2024A&A...689A.191Z}; (ii) the detection of four small glitches in PSR J1048$-$5832 \citep{Zubieta:2022umm, 2024A&A...689A.191Z}; (iii) the observation of the largest glitch ever recorded for PSR J1048$-$5832 \citep{2024ATel16580....1Z}; (iv) the announcement of a glitch in PSR J1740$-$3015 that occurred in late 2022 \citep{2022ATel15838....1Z}; and (v) the confirmation of a glitch in PSR J0742$-$2822 \citep{2022ATel15638....1Z}. We also detected many discrete rotational irregularities in six glitching pulsars different to glitches \citep{2024A&A...689A.191Z}, and a clear change in the pulse profile of PSR J0742$-$2822 after its 2022 giant glitch \citep{2024arXiv241217766Z}.

\section{Data reduction and pulsar timing technique for glitch characterisation}

From the PRESTO package \citep{2011ascl.soft07017R} we used the \textsc{rfifind} task to remove radio-frequency interferences (RFIs) from observations and the \textsc{prepfold} task to fold the observations. We then employed the \textsc{pat} task in the PSRCHIVE package \citep{2004PASA...21..302H} to determine the times of arrival (TOAs) of the pulses with the Fourier phase gradient-matching template fitting method \citep{1992PTRSL.341..117T}. We created the template by using a smoothing wavelet method (\textsc{psrsmooth} task in PSRCHIVE package) to the pulse profile of an observation with a high signal-to-noise ratio that was left out of the posterior timing analysis. 

In order to derive information from the TOAs, we introduced a timing model that aims to predict the TOAs. The discrepancy between the expected and observed TOAs can be used to obtain information about the pulsar rotation.

The temporal evolution of the phase of the pulsar is modelled as a Taylor expansion \citep{2013MNRAS.429..688Y}:

\begin{equation}\label{eq:timing-model}
    \phi(t)=\phi+\nu(t-t_0)+\frac{1}{2}\dot{\nu}(t-t_0)^2+\frac{1}{6}\Ddot{\nu}(t-t_0)^3.
\end{equation}

In Eq.~(\ref{eq:timing-model}), $t_0$ is the reference epoch for the timing model and $\nu$, $\dot\nu$ and $\ddot\nu$ are the rotation frequency of the pulsar and its first and second derivatives at $t_0$. Finally, $\phi$ is the pulse phase at the reference epoch $t_0$. We took from the ATNF pulsar catalogue \citep{ManchesterATNF2005} the initial parameters for the timing model and then we updated them by fitting the model to our TOAs.

During a glitch, the pulsar rotation frequency increases abruptly \citep{2018Natur.556..219P}. This is often accompanied by a decrease in the spin-down rate. In giant Vela glitches, the increase in the rotation frequency can generally be modelled as the sum of one permanent increase in the frequency plus other transitory increases of frequency that decay in their respective scales \citep{Zubieta:2022umm}. 

The additional phase in the pulsar rotation due to the glitch behaviour can be described in the timing model as \citep{glitch-timing}:

\begin{multline}\label{eq:glitch-model}
    \phi_\mathrm{g}(t) = \Delta \phi + \Delta \nu_\mathrm{p} (t-t_\mathrm{g}) + \frac{1}{2} \Delta \dot{\nu}_\mathrm{p} (t-t_\mathrm{g})^2 + \\ 
    \frac{1}{6} \Delta \Ddot{\nu}(t-t_\mathrm{g})^3 + \sum_i \left[1-\exp{\left(-\frac{t-t_\mathrm{g}}{\tau^{i}_\mathrm{d}}\right)} \right]\Delta \nu^{i}_\mathrm{d} \, \tau_\mathrm{d}^{i}.
\end{multline}

Here, $\Delta \phi$ is the offset in the pulsar phase that helps counteract the uncertainty in the glitch epoch $t_g$. The permanent jumps in $\nu$ and its derivatives with respect to the pre-glitch solution are $\Delta \nu_\mathrm{p}$, $\Delta \dot{\nu}_\mathrm{p}$ and $\Delta \Ddot{\nu}$. The transient increments in the frequency are described with $\Delta \nu_\mathrm{d}$, and decay after a timescale $\tau_\mathrm{d}$. For a Vela glitch, the highest number of transient components found so far is four.

The instantaneous changes in $\nu$ and $\dot\nu$ at the glitch epoch can be described as:
\begin{align}
    \Delta \nu_\mathrm{g} &= \Delta \nu_\mathrm{p} + \Delta \nu_\mathrm{d} \\ 
    \Delta \dot{\nu}_\mathrm{g} &= \Delta \dot{\nu}_\mathrm{p} - \frac{\Delta \nu_\mathrm{d}}{\tau_\mathrm{d}} \, .
\end{align}

Then, the degree of recovery of the glitch, $Q$, is characterised by relating the transient and permanent jumps in frequency as $Q=\Delta \nu_\mathrm{d} / \Delta\nu_\mathrm{g}$.

\section{2024 Vela glitch characterisation}\label{sec:J0835}

In \cite{2024ATel16608....1Z} we first reported the detection of a new $(\#23)$ glitch in the Vela pulsar. The 22 glitches previously reported are listed in the
ATNF catalogue {\url{http://www.atnf.csiro.au/people/pulsar/psrcat/glitchTbl.html}}.

On April 28th, we observed the Vela pulsar between MJD 60428.8386 and MJD 60428.9973. We observed $220$ minutes using A1 and $216$ minutes using A2 and measured a barycentric period of $P_\mathrm{bary} = 89.426459082(89)$~ms which was consistent with the ephemeris at that moment. We did not observe a glitch during that observation. Then, we observed the Vela pulsar again on MJD 60431.8304 and measured a period of $P_\mathrm{bary} = 89.426366478(76)$~ms, which shows a decrease with respect to the expected period of $\Delta P = 0.239~\mu$s. We first placed the glitch epoch between MJD 60428.96 (2024-04-28 23h UTC) and MJD 60431.84 (2024-05-01 20h UTC). Later the glitch epoch was constrained to MJD 60429.86962(4) \citep{2024ATel16615....1P}.

In this work, we exhibit a deeper examination of the Vela timing behaviour near the glitch epoch. We focused on a small ($\sim 70~\mathrm{d}$) time window close to the glitch to roughly characterise short-term behaviour of the glitch and avoid the effects of the strong red noise experienced by the Vela pulsar. We included 40 observations made with A1 and 23 observations made with A2. We obtained 4 ToAs ($\sim$1 TOA per hour) from each observation and characterised the white noise using \textsc{TempoNest} \citep{2014MNRAS.437.3004L}. We obtained $TNGlobalEF=4.1283$ and $TNGlobalEQ=-5.64459$. $TNGlobalEF$ is the factor by which the template-fitting underestimates the ToA errorbars, whereas $TNGlobalEQ$ indicates a systematic uncertainty of $2~\mathrm{\mu s}$.

\begin{figure}
    \centering
    \includegraphics[width=\linewidth]{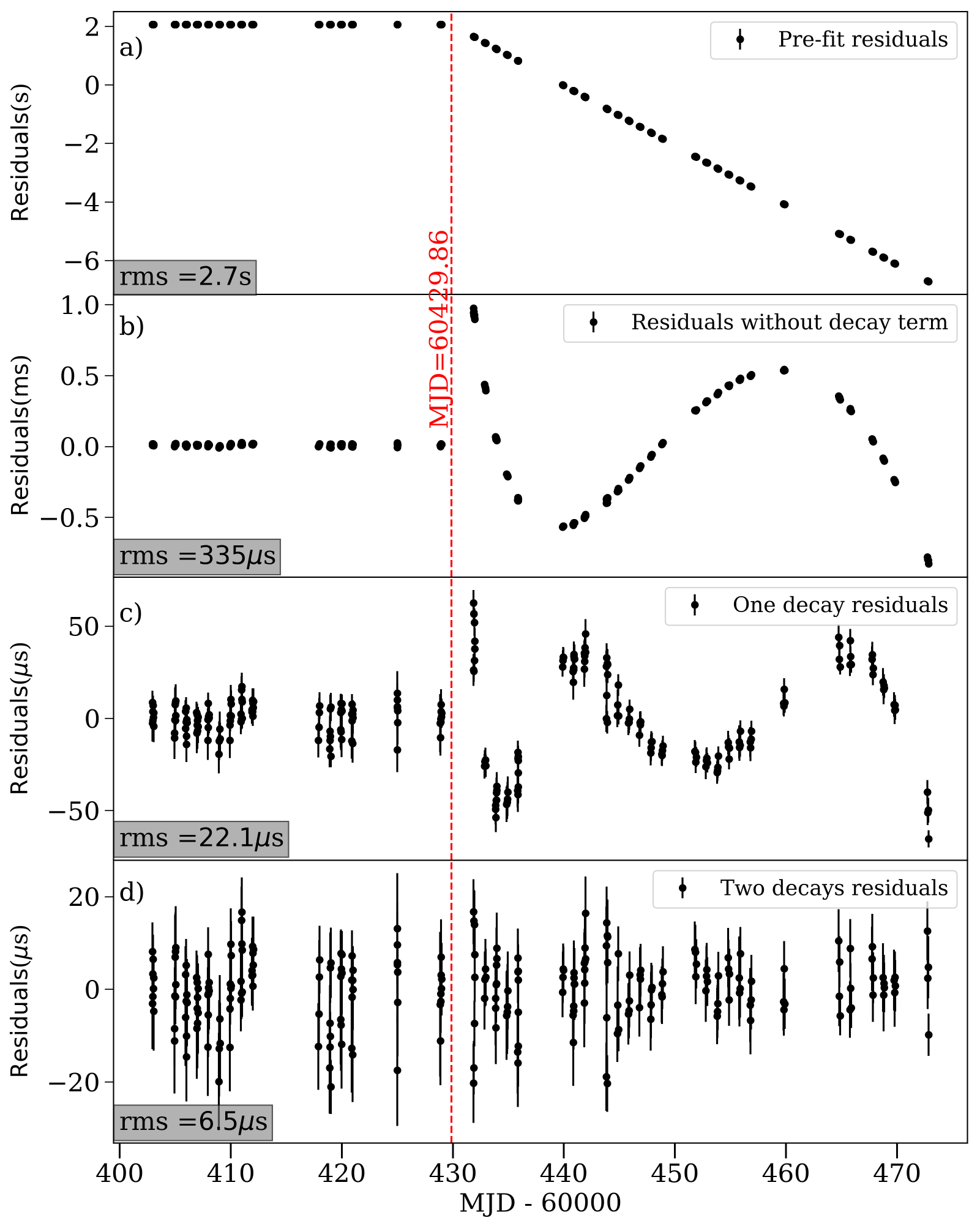}
    \caption{Vela's timing model with the parameters from Table~\ref{tab:Vglitch}.} 
    \label{fig:VelaResiduals}
\end{figure}

In order to obtain a full-timing solution for the glitch, we followed a procedure similar to what we did in \cite{Zubieta:2022umm}. We first obtained the pre-glitch rotational model by fitting $\nu$, $\dot\nu$ and $\ddot\nu$ to the TOAs before the glitch, keeping the value of $DM=67.93(1)~\mathrm{pc\,cm^{-3}}$ from the ATNF pulsar catalogue\footnote{\url{http://www.atnf.csiro.au/people/pulsar/psrcat/}}. We show the residuals corresponding to the pre-glitch timing model in Fig.~\ref{fig:VelaResiduals}a). We then fit to the TOAs the permanent jumps in $\nu$ and $\dot \nu$ ($\Delta\nu$ and $\Delta\dot\nu$) in Eq.~(\ref{eq:glitch-model}). At this stage we also included $\Delta \phi$ in the timing model, assuming $t_g=60429.86962(4)$ as reported by \cite{2024ATel16615....1P}. 

The timing residuals shown in Fig.~\ref{fig:VelaResiduals}b) revealed the presence of a recovery term. We used the \textsc{glitch} plug-in in TEMPO2 to estimate $\tau_{d1}\sim17\mathrm{d}$. We included $\tau_{d1}$ in the timing model together with $\nu_{d1}$ and fitted the whole glitch model to the residuals. We still found a remaining structure in the residuals as shown in Fig.~\ref{fig:VelaResiduals}c). We ran the \textsc{glitch} plug-in again over these residuals and we found one extra recovery term of $\sim 2\mathrm{d}$.

\begin{figure}
    \centering
    \includegraphics[width=\linewidth]{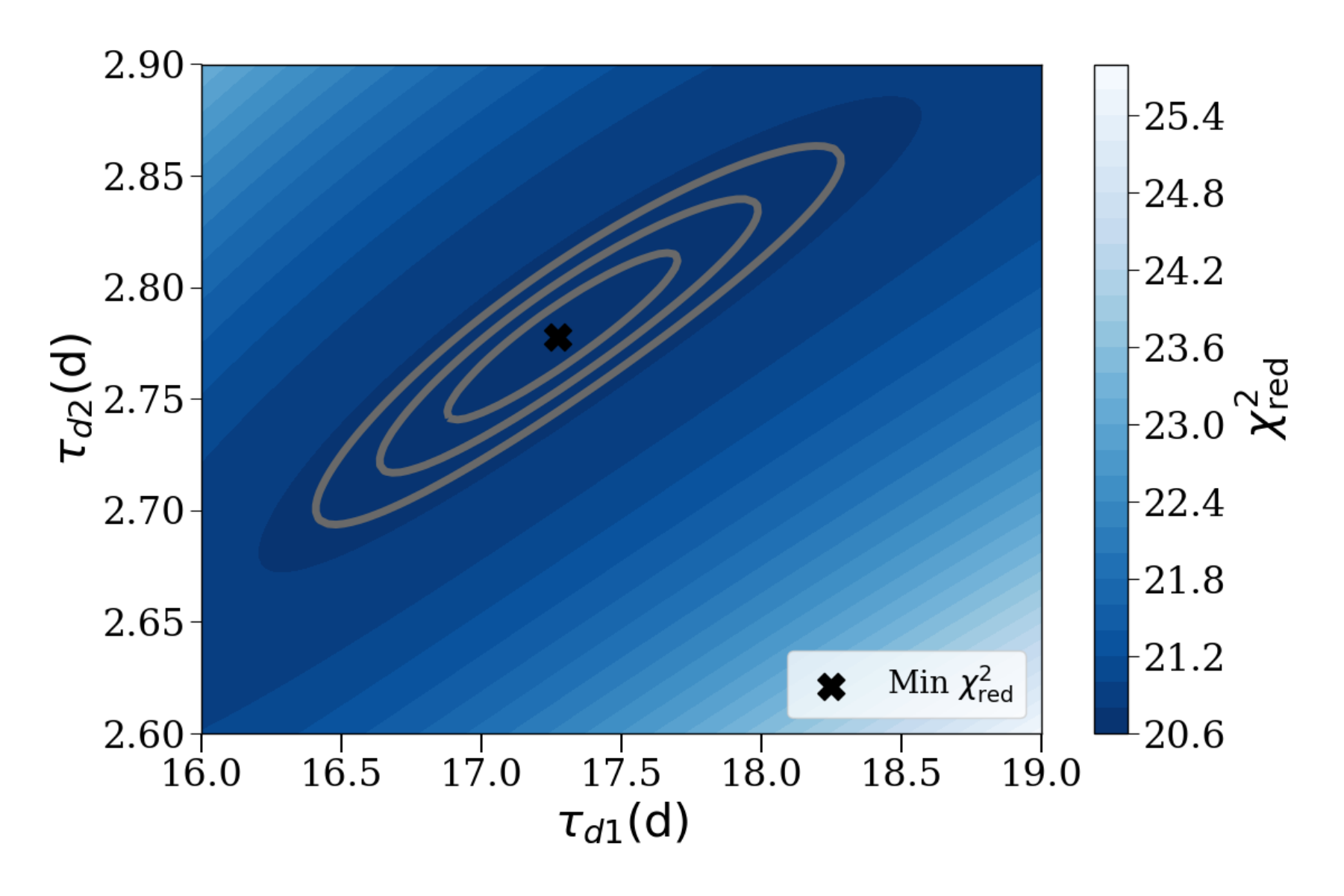}
    \caption{Best fit of the decay time constants $\tau_{d1}$ and $\tau_{d2}$ for the 2024 Vela glitch. The solid line, dashed line, and dot-dashed line indicate the 1-, 2- and 3-$\sigma$ confidence regions.}
    \label{fig:VelaTau}
\end{figure}

Considering the difficulty of TEMPO2 to fit recovery terms to the TOAs, we looked for the best combination of $\tau_{d1}$ and $\tau_{d2}$ as we did in \cite{Zubieta:2022umm}. We defined ranges for $\tau_{d1}$ and $\tau_{d2}$, and fitted $\Delta \phi$, $\Delta \nu$, $\Delta \dot \nu$, $\Delta \nu_{d1}$ and $\Delta \nu_{d2}$ for every possible combination of $\tau_{d1}$ and $\tau_{d2}$ in those ranges. We searched for the glitch timing model that minimised the reduced chi-squared of the timing residuals ($\chi_{red}^2=\chi^2/\mathrm{dof}$), where $\mathrm{dof}$ are the degrees of freedom. We then systematically shortened the ranges for $\tau_{d1}$ and $\tau_{d2}$ and arrived to the solution that we show in Fig. \ref{fig:VelaTau}, where we calculated the $1\sigma$, $2\sigma$ and $3\sigma$ errorbars following \cite{Press1992}.
We obtained $\tau_\mathrm{d1}=17.3(3)~\mathrm{d}$ and $\tau_\mathrm{d2}=2.78(3)~\mathrm{d}$, with a degree of recovery of $Q_1=0.562(2)\%$ and $Q_2=0.462(1)\%$ respectively. In this case, the degree of recovery for the first and second recovery scale are similar, and the sum of both recoveries is around $1\%$ of the total glitch size.

We plotted the residuals with the best solution in Fig.~\ref{fig:VelaResiduals}d). Residuals are flat indicating that for our data the glitch timing model is complete. We did not find evidence of a step change in $\Ddot\nu$ during the glitch. The full timing model is shown in Table~\ref{tab:Vglitch}.

\begin{table}
  \centering    
  \caption{Parameters of the timing model for the 2024 April 29th Vela glitch and their 1$\sigma$ uncertainties.}
   \begin{tabular}{ll}        
     \hline
     Parameter & Value \\
     \hline
     $\mathrm{PEPOCH}$ (MJD)& 60408 \\
     $\mathrm{F0}(\mathrm{s^{-1}})$& 11.18285953(1)\\
     $\mathrm{F1}(\mathrm{s^{-2}})$ &  $-1.55405(4)\times 10^{-11}$\\
     $\mathrm{F2}(\mathrm{s^{-3}})$ &  $8.4(2)\times 10^{-22}$\\
     $\mathrm{DM}(\mathrm{cm^{-3} pc})$ & 67.93(1)\\
     $t_\mathrm{g}$ (MJD) &60429.86961(4)\\
     $\Delta\nu_\mathrm{p}$ (s$^{-1}$) & $2.65752(3) \times 10^{-5}$\\
     $\Delta\dot{\nu}_\mathrm{p}$ (s$^{-2}$)& $-1.0140(8)\times 10^{-13}$\\
     $\Delta\ddot{\nu}$ (s$^{-3}$)& - \\
     $\Delta\nu_\mathrm{d1}$\, (s$^{-1}$)&$1.510(4)\times 10^{-7}$\\
     $\tau_\mathrm{d1}$ (days) & 17.3(3)\\
     $\Delta\nu_\mathrm{d2}$\, (s$^{-1}$)&$1.242(3)\times 10^{-7}$\\
     $\tau_\mathrm{d2}$ (days) & 2.78(3)\\
     $\Delta \phi$ & 0.00676(9)\\
     $\Delta\nu_\mathrm{g}/\nu$ & $2.40103(5)\times 10^{-6}$\\
     $\Delta\dot\nu_\mathrm{g}/\dot\nu$ & 0.107(1)\\
     $Q_1$ & 0.00562(2)\\
     $Q_2$ & 0.00462(1)\\
     $TNGlobalEF$ & 4.1283 \\
     $TNGlobalEQ$ & -5.64459 \\
     \hline
   \end{tabular}
  \label{tab:Vglitch}
 \end{table}

\begin{figure}
    \centering
    \includegraphics[width=\linewidth]{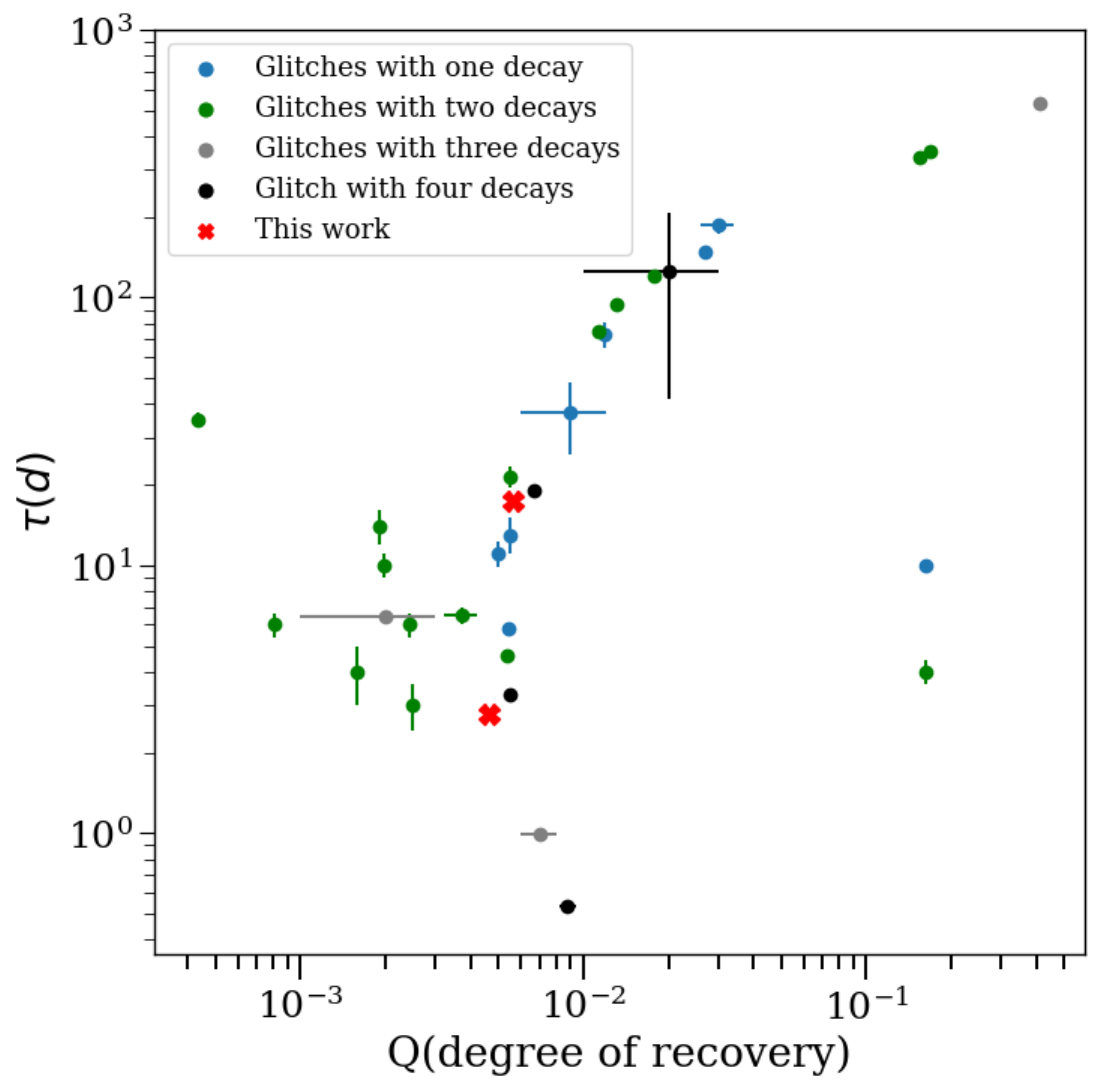}
    \caption{Comparison of current and previous glitches decaying parameters for Vela pulsar.}
    \label{fig:Qtau}
\end{figure}

In \cite{2024A&A...689A.191Z} we reported a third large recovery term for the 2021 Vela glitch. Together with the two decays terms that we report in this work for the 2024 Vela glitch, we update in Fig. \ref{fig:Qtau} all the recovery terms reported so far for Vela glitches. It is interesting to note that the third recovery time scale that we reported in \cite{2024A&A...689A.191Z} for the 2021 glitch, is not only the largest time scale reported so far but also it has the highest degree of recovery. This remarks the importance of re-analysing glitches with the whole post-glitch data span (until the following glitch arrives). 

In addition, Fig. \ref{fig:Qtau} shows that, except for two data points, the degree of recovery appears to increase with the recovery timescale.  It is important to consider that there is an observational bias (and degeneracy) regarding both the number of decaying components being fitted and the corresponding timescales obtained \citep{2022RPPh...85l6901A}. Therefore, it is important to keep revising glitch solutions to better constrain this effect and gain insight into the behaviour of the neutron star interior. For example, the need to fit multiple exponential components with different timescales, in terms of the vortex creep model, suggests that different regions within the neutron star respond to the glitch according to their own intrinsic properties \citep{1993ApJ...409..345A,2020MNRAS.496.2506G}. By improving our understanding of these timescales, we could better interpret how the spatial distribution of the pinning forces affects the overall relaxation process.

\section{Analysis Methods: Pulse-by-pulse analysis of the 2021 Vela glitch}\label{sec:analysis}

In this Section, we report the analysis of the observations around the Vela glitch pulse by pulse. 
High-resolution single-pulse micro-structure pulse studies of the Vela pulsar were reported in \cite{Johnston+2001} and \cite{Cairns+2001}, while the temporal evolution of the pulses for large timescales was studied in \cite{Palfreyman+2016}.
Here we take advantage of the large amount of our daily data, which is well suited for statistical and machine learning studies. Our approach has been carried out using a combination of the VAE reconstruction and the SOM clustering techniques \citep{kingma2014autoencoding,teuvo1988som}. 

\begin{figure*}[h!]
   \centering 
   \includegraphics[width=\textwidth]{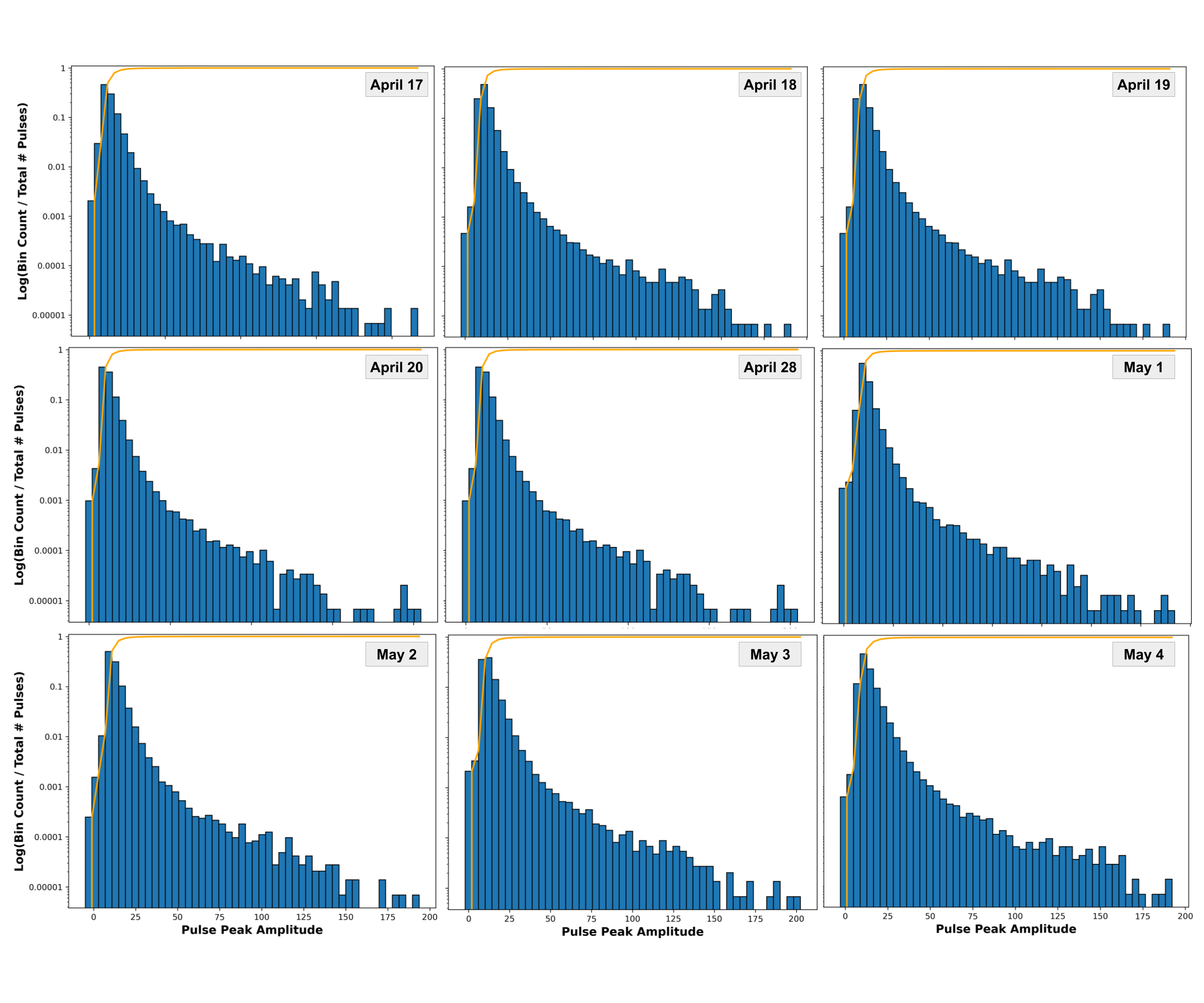}
     \caption{
     Peak amplitude of single pulses distribution for observations with A2 preglitch on April 17-20,28, and post glitch on May1-5, 2024. The top yellow curve is the cumulative sum.}
  \label{fig:17New}
 \end{figure*}

{We analysed nine observations on 2024, five before the glitch, on April 17th, 18th, 19th, 20th, and 28th, and four after the glitch, on May 1st, 2nd, 3rd, and 4th. All observations were carried out with A2 in dual polarisation in the 1400--1456~MHz band. 
The number of pulses in each observation is given in Table~\ref{tab:observations}. Those are uninterrupted single observations with A2 antenna observations typically lasting 3.66 hours.
All observations were folded with a fixed  $DM=67.93(1)$~pc\,cm$^{-3}$  from the ATNF catalogue\footnote{\url{http://www.atnf.csiro.au/people/pulsar/psrcat/}} (as we have seen very small variations during each observation, $DM<0.2$~pc\,cm$^{-3}$)
and cleaned from radio frequency interferences using \texttt{RFIClean} \citep{Maan2020} with protection of the fundamental frequency of Vela (11.184~Hz). The complete procedure is described in Appendix C of \cite{Lousto:2021dia},
where we found that using \texttt{rfifind} \citep[a task within \texttt{PRESTO};][]{PRESTO} on the data output from \texttt{RFIClean} further improves the S/N in most of the cases we studied. The amplitudes of the pulses are in arbitrary units as we did not observe any flux calibrator. Their relative distribution, day per day analysed here (Table~\ref{tab:observations}), is displayed in Fig.~\ref{fig:17New}. We note the qualitative similarities of the A2 pulse distributions pre-glitch on top, while the post-glitch observations are a bit more heterogeneous. For a more quantitative comparison, one can look at the parameters of the clusters in the Tables in Appendix~\ref{sec:appendix}.}

\begin{table*}
	\centering
	\caption{
 Date of each observation with A2, duration in hours, the MJD at the topocentric beginning of the observations, the corresponding number of single pulses analysed, instantaneous topocentric period, $P_{\mathrm{obs}}$, and estimated signal to noise ratio (SNR) for the selected observations around the 2024 Vela glitch used for the pulse-by-pulse analysis. The estimated time of the glitch on April 29 is MJD 60429.8696 .}
	\label{tab:observations}
	\begin{tabular}{l c c c l l l} 
		\hline
		Date & Antenna & Duration [h] & epoch MJD & \# pulses & $P_{\mathrm{obs}}$ [ms] & SNR\\
		\hline
  April 17 & A2 & 3.66 & 60417.86861037355 & 147394 & 89.42592924 & 348.18 \\
  April 18 & A2 & 3.66&60418.86988352169 &147394 &89.42598438 & 274.62 \\
  April 19 & A2 & 3.66&60419.86687426244 &147363 &89.42603628 & 263.84  \\
  April 20 & A2 & 3.66&60420.61084910590  &147389 &89.42608723 & 261.88   \\
  April 28 & A2 & 3.60&60428.34752565132 &144793 &89.42645908 & 219.39  \\
  \hline
  May 1 & A2 & 3.60&60431.83923799470 &144791 &89.42636648 &276.11   \\
  May 2 & A2 & 3.66&6043283136500318 &147394 &89.42640220 & 256.08  \\
  May 3 & A2 & 3.66&60433.82876083651 &147393 &89.42643953 &326.08   \\
  May 4 & A2 & 3.53 &60434.82621454021&142092 &89.42647504 &  300.34 \\
  \hline
	\end{tabular}
\end{table*}

\subsection{Self-Organizing Map (SOM) techniques}\label{sec:CS}
In Refs.~\cite{Lousto:2021dia,Zubieta:2022umm} we described a deep learning generative and clustering method built on Variational AutoEncoders (VAE) and Self-Organizing Maps (SOM) to perform Vela per-pulse clustering in an unsupervised manner. 

For our method, we employ a two-stage process where the raw noisy pulses are first de-noised (VAE) and then are grouped into clusters second (SOM). The raw noisy pulses $\mathbf{X}$ are denoised into smooth approximations $\mathbf{\hat{X}}$ through neural networks that compress the input into a lower-dimensional stochastic space and then try to reconstruct the signal. We then define a 2D grid of $M$ nodes, $\mathbf{V}_{1:M}$, each initialised as a random vector in data space. The grid is iteratively updated through a competitive process where the input signals are presented to all nodes and the closest node via a distance measure (e.g. Euclidean distance) is chosen as the 'best matching unit'. This node and its grid neighbours are then slightly pulled closer to that input data point. This process is repeated until the grid is stable. The result is a set of cluster centres and assignments that partition similar signals into groups based on the dataset's latent structure.
The schematic diagrams of VAE and the usage of SOM for clustering are presented in Fig.~11 of \cite{Lousto:2021dia}.

\subsection{Results}\label{sec:results}

\begin{figure*}[h!]
   \centering 
   \includegraphics[width=\textwidth]{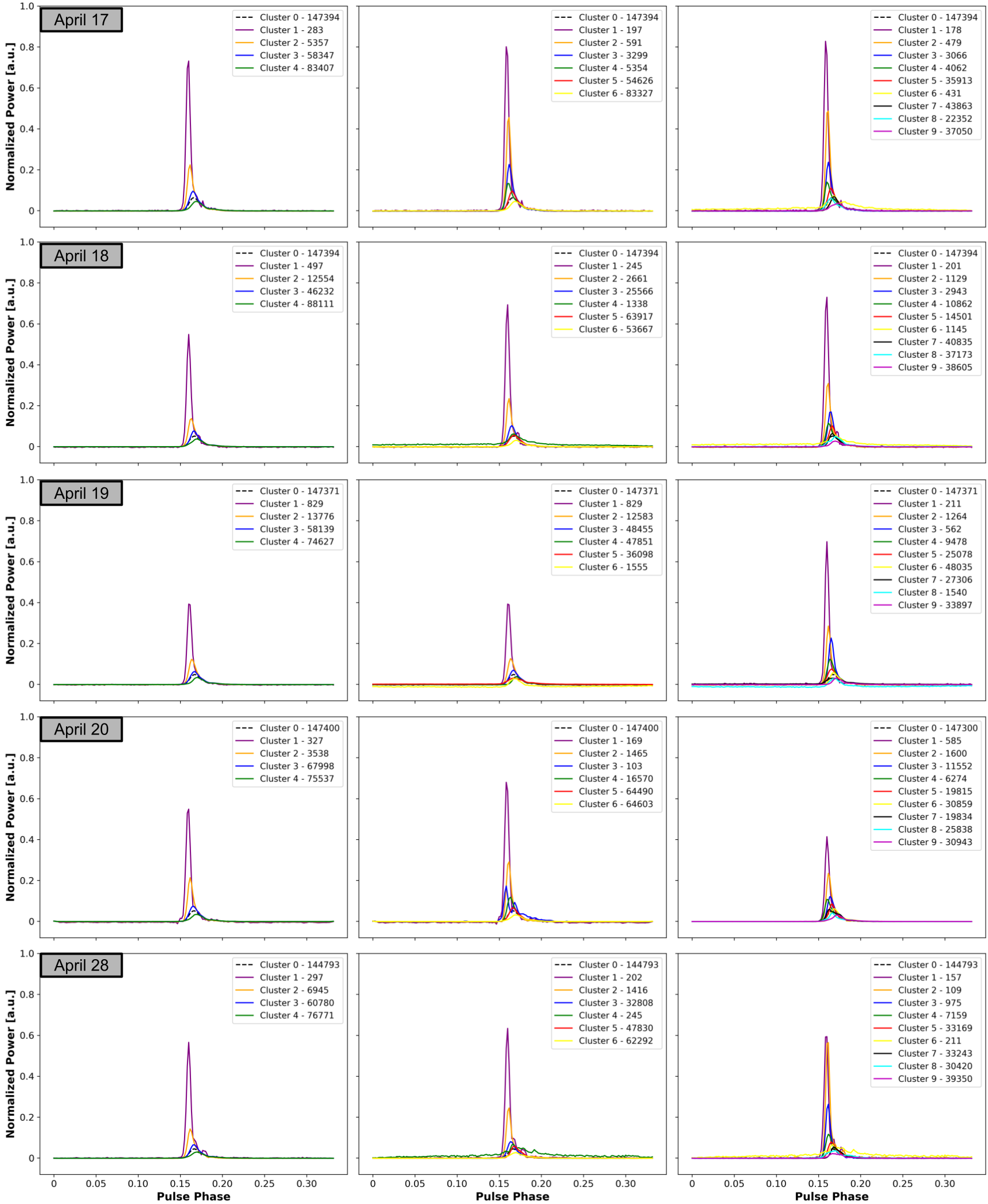}
     \caption{
     Mean cluster reconstruction for observations with A2 before the glitch on 2024, April 17, 18, 19, 20, and 28, using 4, 6, and 9 SOM clustering. [200 (out of total 611) phase bins were taken around the mean peak (at bin 100) of each day to perform the single-pulse analysis].}
  \label{fig:Allbg}
 \end{figure*}

\begin{figure*}[h!]
   \centering 
   \includegraphics[width=\textwidth]{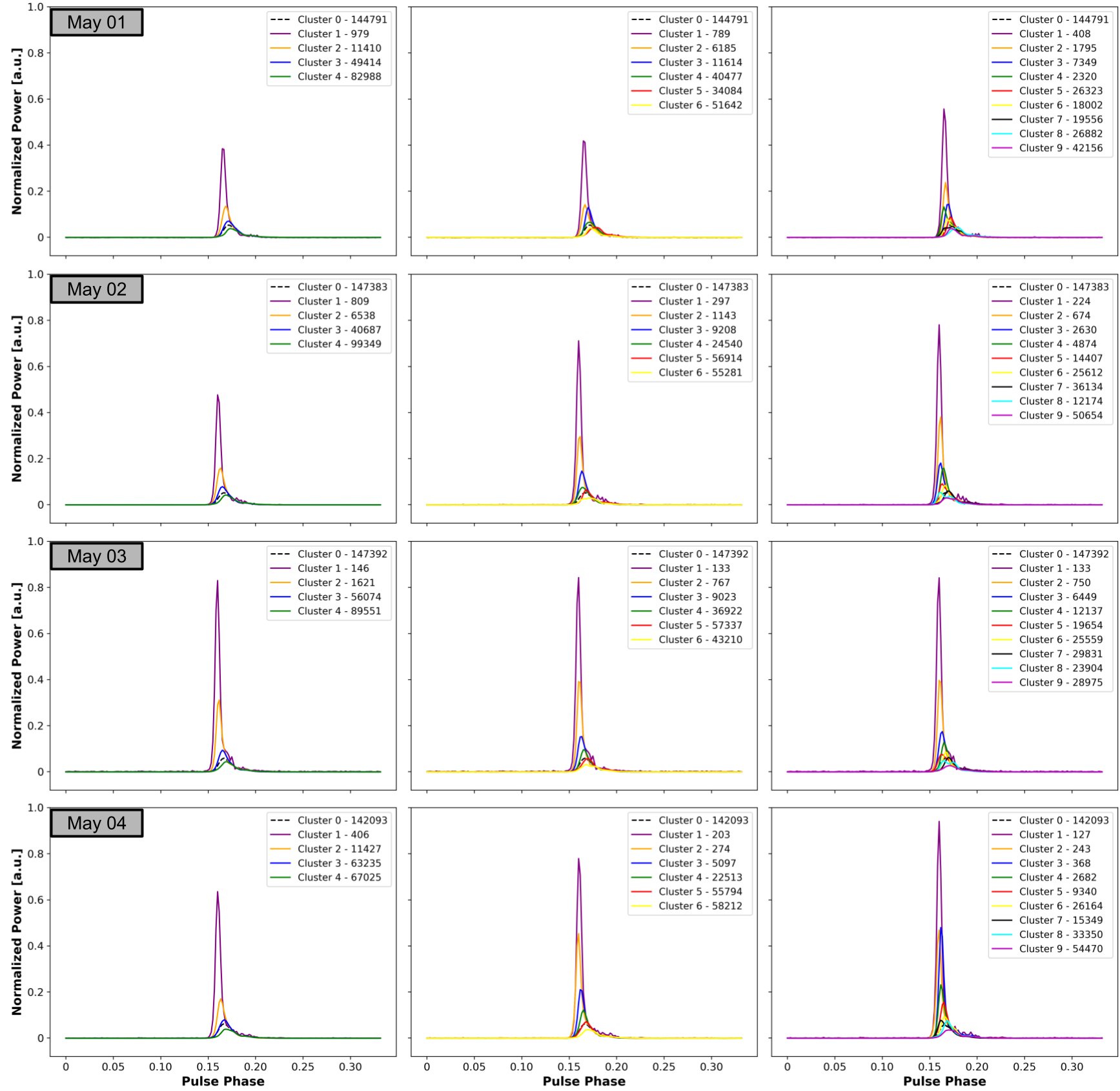}
     \caption{
     Mean cluster reconstruction for observations with A2 after the glitch on 2024, May 1, 2, 3, and 4, using 4, 6, and 9 SOM clustering. [200 (out of total 611) phase bins were taken around the mean peak (at bin 100) of each day to perform the single-pulse analysis].}
  \label{fig:Allag}
 \end{figure*}

We have collected the results of the SOM clustering for the nine days of observation in Figs.~\ref{fig:Allbg}
and \ref{fig:Allag}.
The results are displayed by days in successive rows and the three columns correspond to the choice of collecting the whole set of pulses in 4, 6, and 9 clusters respectively. The glitch, which occurred on 2024 April 29th, would lie between those figures. 
We have chosen the same vertical scale to represent the mean pulse of each cluster over the choices of the number of clusters and over the days of observation in order to exhibit the relative amplitudes, also affected by the different amounts of observing time.
Figures~\ref{fig:Allbg}-\ref{fig:Allag},  display pulses amplitudes (in the arbitrary units
coming from the \texttt{PRESTO} (FFT) normalisation). We have not used standard sources to seek a normalisation of the observations, although we report 
the signal-to-noise ratio (SNR) of each observation as provided by \texttt{PRESTO} in Table~\ref{tab:observations}.

The labelling of the clusters in each panel is ordered from the largest to the lowest amplitude mean pulse, while cluster 0 is the total mean pulse of the whole observation and remains the same over the three horizontal panels as a reference value.
We first note an increase in the amplitude of the mean pulse of the cluster 1 as we increase the number of clusters allowed to SOM. They also decrease the number of pulses per cluster (as expected), which explains the amplitude increase. This behaviour is shared by clusters 2 and 3, and successively. We also note an earlier arrival and a mild decrease in the width of the high amplitude clusters (feature that could be used for improved timing in other circumstances or for other millisecond pulsars as we noted in \cite{Lousto:2021dia}).
These points are more precisely quantified, with estimated errors, in the Tables presented in Appendix~\ref{sec:appendix}.

On the other hand, we have not found any notable qualitative systematic changes in the clustering during the days just before and after the glitch.
We hence conclude that the effects of the glitch on the pulses might be either subtler and / or of shorter time scale around the glitch time and we will continue to monitor Vela to capture the next event 'live' as in the case of the 2016 glitch \citep{2019NatAs...3.1143A}.
 
\section{Conclusions and discussion}\label{sec:conclusions}

\begin{figure}[h!]
	\includegraphics[width=\columnwidth]{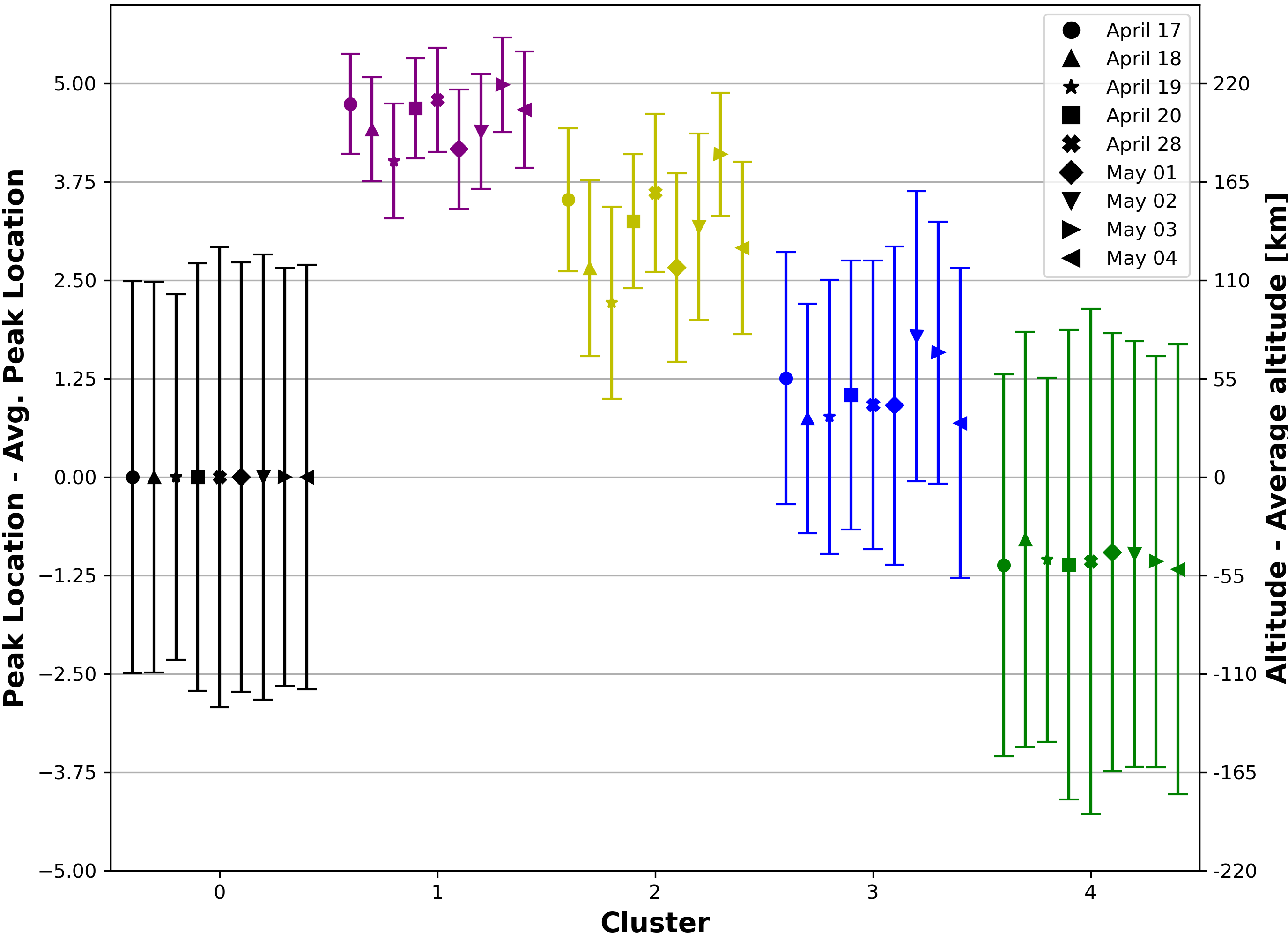}
    \caption{Peak location and magnetosphere altitude, with the corresponding error bars, for each of the pulse 4-clusters and the whole observation.}
    \label{fig:altitudes}
\end{figure}

Here we presented a detailed analysis of the latest $(\#23$ recorded) 
large Vela glitch, first reported in \cite{2024ATel16608....1Z}, 
finding for this 2024 glitch a
$(\Delta\nu_\mathrm{g}/\nu)_{2024}=2.4\times10^{-6}$, 
comparable with the values for the previous 2019 and 2021 glitches. 
We are able to provide an accurate description of the glitch characteristic epoch, jumps, and exponential recoveries with timescales of 17.3 and 2.78 d, respectively,
(See Table~\ref{tab:Vglitch} and Fig.~\ref{fig:Qtau}). 
In 2019 we reported a large Vela $(\#21$ recorded)
glitch \citep{atel_vela,Gancio2020} with a $(\Delta\nu_\mathrm{g}/\nu)_{2019}=2.7\times10^{-6}$. More recently, in \cite{2021ATel14806....1S,Zubieta:2022umm}
we presented a detailed analysis of the $(\#22$ recorded)
2021 Vela glitch, with $(\Delta\nu_\mathrm{g}/\nu)_{2021}=1.2\times10^{-6}$ and exponential recoveries with associated timescales of 6.4 and 1 d, respectively.

The high cadence of our observations allowed us to verify and independently estimate the time of the glitch as, 
$t_g(\text{MJD})=60429.86961(4)$,
confirming the value initially reported in \cite{2024ATel16615....1P}. Furthermore, the accuracy of our observations also allowed us to perform pulse-by-pulse studies of Vela using the machine learning techniques previously validated in \cite{Lousto:2021dia,Zubieta:2022umm}.

For the sake of direct comparison among Vela's last two large glitches we have paralleled the pulse-by-pulse analysis of this 2024 Vela glitch with that of the 2021 in \cite{Lousto:2021dia,Zubieta:2022umm}.
We note in Figs.~\ref{fig:Allbg}--\ref{fig:Allag} that the higher amplitude pulse clusters tend to appear earlier and are narrower than the bulk of the other pulses (see also tables in Appendix~\ref{sec:appendix}). This is particularly rigorous cluster by cluster in the four cluster analysis and typically large amplitude clusters are narrower by a factor $\sim 2$ and the errors are nearly an order of magnitude smaller than for the whole set of pulses.
We note that these four-cluster distributions follow a similar pattern in the sense to our previous studies with observations about six months before the 2021 glitch, on 2021 January 21, 24, 28 and March 29 \citep{Lousto:2021dia} and around the 2021 glitch on July 22nd 2021 and on July, 19th, 20th, 21st, 23rd,
and 24th \citep{Zubieta:2022umm}; and it was associated with strata of the magnetosphere at different highs separated by $\sim 100$~km \citep{Lousto:2021dia}. Fig.\ref{fig:altitudes} displays the results of applying this model to each of the four days of observation for each antenna. 
The right hand side ordinate gives
the components distances to the average pulse reference height in the pulsar magnetosphere. We note the consistency between the components for each of the four days and for each individual antenna's observations. The four components appear to be almost equidistant (this maybe an effect of the SOM clustering method) and roughly of the order of $\sim100$ kilometers.
We also note that new independent recent studies confirm this early arrival of high amplitude pulses \cite{Mahida:2023maa}.

We finally note the much smaller error bars displayed in Fig. \ref{fig:altitudes} in the pulse peak determination for the larger amplitude pulsar cluster (labelled as \#1) than for the whole observation (labelled as \#0), thus opening the possibility to use this cluster for timing of pulsars in order to achieve  much higher precision of its timing measurements.

\section*{Acknowledgements}


We especially thank Yogesh Maan for numerous beneficial discussions about the best use of \texttt{RFIClean}. 
COL gratefully acknowledge the National Science Foundation (NSF) for financial support from Grants No.\ PHY-1912632, and PHY-2207920. JAC and FG are CONICET researchers. JAC and FG acknowledge support from PIP 0113 (CONICET). JAC and FG were also supported by grant PID2019-105510GB-C32/AEI/10.13039/501100011033 from the Agencia Estatal de Investigaci\'on of the Spanish Ministerio de Ciencia, Innovaci\'on y Universidades, and by Consejer\'{\i}a de Econom\'{\i}a, Innovaci\'on, Ciencia y Empleo of Junta de Andaluc\'{\i}a as research group FQM-322, as well as FEDER funds. FG acknowledges support from PIBAA 1275 (CONICET). 





\bibliographystyle{aa}
\bibliography{biblio,bibliografia,references} 




\appendix

\section{Tables of SOM Clustering}\label{sec:appendix}

Here we include the numerical information in tabular form about the 
clustering analysis summarised in Fig.~\ref{fig:Allbg}-\ref{fig:Allag}. They include a 6 SOM clusters decomposition as representative for each of the days of observation.
We provide the number of pulses of each cluster \# pulses; peak location from the index of the maximum value in the pulse sequence;
peak height from the maximum value of the pulse sequence;
peak width done by first finding the maximum value of the sequence, then performing full-width half maximum of peak;
(library used for this: \url{https://docs.scipy.org/doc/scipy/reference/generated/scipy.signal.peak_widths.html});
for the peak skew we evaluated the Fisher-Pearson coefficient of skewness; 
(using the scipy for this computation \url{https://docs.scipy.org/doc/scipy/reference/generated/scipy.stats.skew.html});
The cluster \#0 corresponds to the total number of pulses in the observation and the successive clusters from \#1 to the \#6 SOM clustering are ordered accordingly to the highest peak amplitude of the mean pulse computed for each cluster and represented in Figs.~\ref{fig:Allbg}-\ref{fig:Allag}.
We compute the peak location with respect to our grid of bins (here centred at around 100 for cluster \#0) and totalling 611 bins per period, giving us a time resolution of 146~$\mu$s. We also provide a measure of the pulse width as given by the standard deviation $(\sigma)$ and its skewness, all with estimated 1-$\sigma$ errors,
and finally MSE is the standard mean squared error $\sum_{i=1}^N(x_i-\bar{x})^2/N$, the average per-step mean squared reconstruction error over all sequences.
We observe a systematic tendency for the pulses' peaks to appear earlier the higher the amplitude as well as a reduction of its width and an increase of the skew (also observed in the previous work of \cite{Lousto:2021dia,Zubieta:2022umm} analysing  2021 observations.

In Tables~A.1.--A.9. we describe in detail the six-cluster analysis of the observations on the days
\input{figures/statistics_as_latex.txt},
\input{figures/statistics_as_latex-2.txt},
\input{figures/statistics_as_latex-3.txt},
\input{figures/statistics_as_latex-4.txt},
\input{figures/statistics_as_latex-5.txt},
\input{figures/statistics_as_latex-6.txt},
\input{figures/statistics_as_latex-7.txt},
\input{figures/statistics_as_latex-8.txt}, and
\input{figures/statistics_as_latex-9.txt}.

\section{VAE reconstruction and SOM Clustering for April 28 observation with A2}\label{sec:appendix2}

\begin{figure*}
    \centering 
      \includegraphics[width=0.85\textwidth]{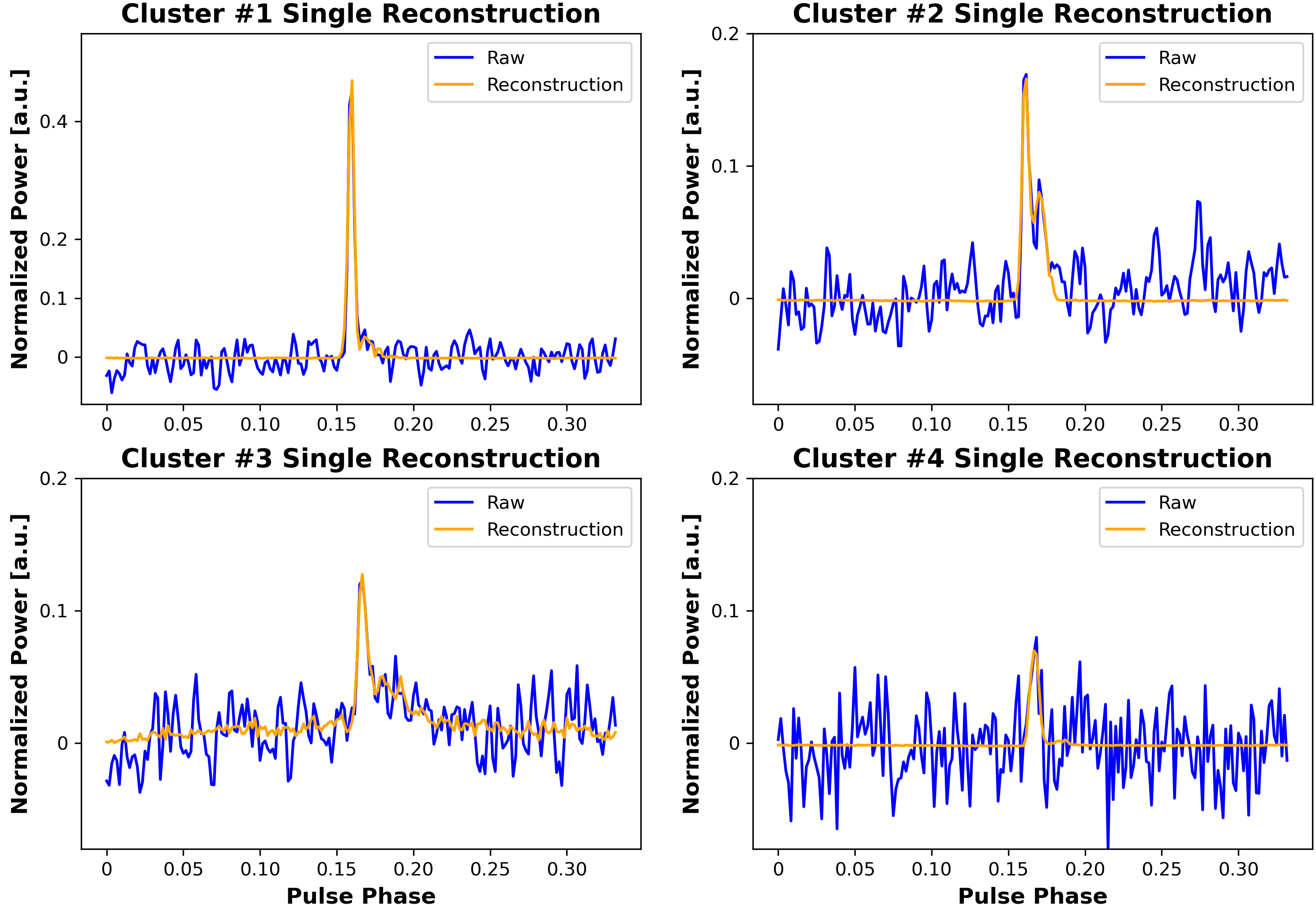}
      \caption{Sample of VAE pulse reconstruction for the April 28, 2024, observations with A2 for 4 SOM clustering. Pulses \# 47419, 36415, 35083, 53075, respectively.}
   \label{fig:7-20pulses}
  \end{figure*}

\begin{figure*}[h!]
    \centering 
      \includegraphics[width=0.85\textwidth]{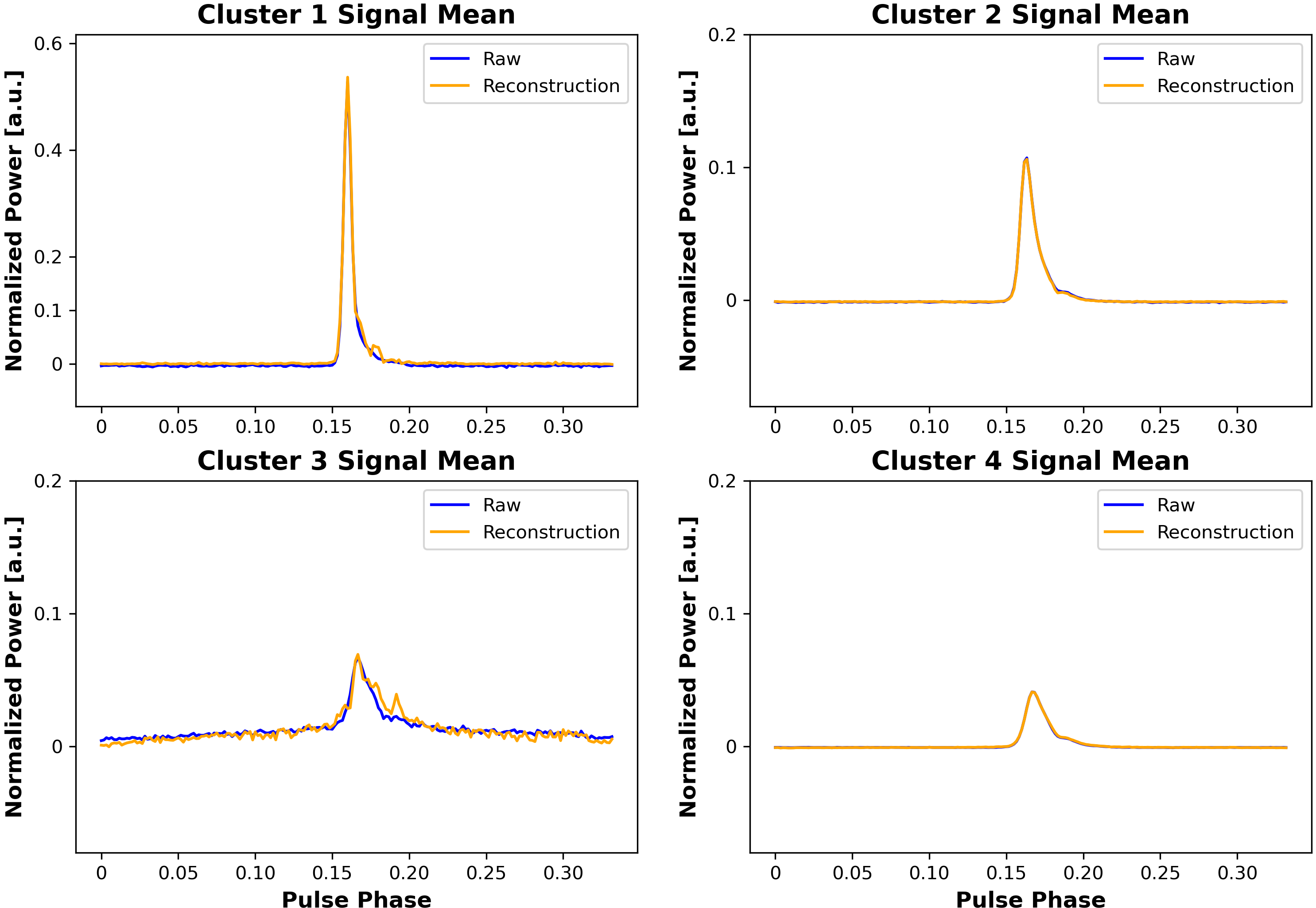}
      \caption{VAE reconstruction of the 4 SOM clustering for the April 28, 2024, observations with A2.}
   \label{fig:meanSigComp}
  \end{figure*}

In order to show that what we observe with the clusters baseline is not an artifact of the VAE pulse reconstruction method,
in Fig.~\ref{fig:7-20pulses} we display some selected {\it individual} raw pulses belonging to the 4 SOM clusters versus their corresponding reconstructions showing the actual baseline fluctuations over the full period range. We also display the VAE representation of the four SOM clusters in Fig.~\ref{fig:meanSigComp}.

\end{document}

%% file: figures/statistics_as_latex.txt
april-17-2024-A2
\begin{table*}
	 \centering
	 \caption{SOM Clustering for April 17 with Antenna A2.}
	 \label{tab:[]}
	 \begin{tabular}{cclllll}
	 	 \hline
	 	 Cluster \# & \# Pulses & Peak Loc & Peak Height & Peak Width & Peak Skew & MSE \\ 
	 	 \hline 
	 	 0 & 147394 & $100.37 \pm 2.49$ & $11.14 \pm 7.67$ & $6.40 \pm 2.59$ & $4.57 \pm 0.90$ & $0.00005 \pm 0.00008$  \\ 
	 	 1 & 215 & $95.49 \pm 0.57$ & $108.43 \pm 31.96$ & $3.14 \pm 0.32$ & $7.16 \pm 0.43$ & $0.04965 \pm 0.32456$  \\ 
	 	 2 & 1187 & $97.11 \pm 0.80$ & $49.57 \pm 16.37$ & $3.25 \pm 0.49$ & $6.22 \pm 0.77$ & $0.00659 \pm 0.00999$  \\ 
	 	 3 & 3055 & $96.48 \pm 0.84$ & $28.17 \pm 7.74$ & $3.86 \pm 0.93$ & $5.58 \pm 0.81$ & $0.00246 \pm 0.00379$  \\ 
	 	 4 & 29349 & $98.71 \pm 1.48$ & $16.73 \pm 4.45$ & $4.36 \pm 1.04$ & $4.78 \pm 0.70$ & $0.00025 \pm 0.00038$  \\ 
	 	 5 & 43484 & $99.48 \pm 1.65$ & $10.70 \pm 2.64$ & $5.33 \pm 1.82$ & $4.54 \pm 0.83$ & $0.00016 \pm 0.00025$  \\ 
	 	 6 & 70104 & $101.87 \pm 2.37$ & $7.38 \pm 2.60$ & $6.51 \pm 2.80$ & $4.42 \pm 0.93$ & $0.00010 \pm 0.00015$  \\ 
	 	 \hline 
	 \end{tabular} 
\end{table*} 

%% file: figures/statistics_as_latex-2.txt
april-18-2024-A2
\begin{table*}
	 \centering
	 \caption{SOM Clustering for April 18 with Antenna A2.}
	 \label{tab:[]}
	 \begin{tabular}{cclllll}
	 	 \hline
	 	 Cluster \# & \# Pulses & Peak Loc & Peak Height & Peak Width & Peak Skew & MSE \\ 
	 	 \hline 
	 	 0 & 147394 & $100.45 \pm 2.48$ & $9.29 \pm 6.52$ & $5.43 \pm 1.93$ & $4.50 \pm 0.89$ & $0.00005 \pm 0.00007$  \\ 
	 	 1 & 261 & $95.84 \pm 0.64$ & $94.11 \pm 28.18$ & $3.25 \pm 0.24$ & $7.05 \pm 0.49$ & $0.03725 \pm 0.18165$  \\ 
	 	 2 & 3275 & $96.87 \pm 0.79$ & $30.46 \pm 10.81$ & $3.44 \pm 0.53$ & $5.84 \pm 0.71$ & $0.00213 \pm 0.00321$  \\ 
	 	 3 & 25040 & $98.86 \pm 1.28$ & $14.64 \pm 3.68$ & $4.51 \pm 1.21$ & $4.98 \pm 0.70$ & $0.00027 \pm 0.00040$  \\ 
	 	 4 & 1628 & $100.65 \pm 2.29$ & $9.62 \pm 3.26$ & $4.96 \pm 1.56$ & $3.51 \pm 0.91$ & $0.00356 \pm 0.00529$  \\ 
	 	 5 & 63484 & $100.29 \pm 2.11$ & $8.89 \pm 2.15$ & $5.35 \pm 1.98$ & $4.46 \pm 0.71$ & $0.00010 \pm 0.00016$  \\ 
	 	 6 & 53706 & $101.63 \pm 2.67$ & $5.56 \pm 1.93$ & $3.72 \pm 0.90$ & $4.25 \pm 0.96$ & $0.00012 \pm 0.00018$  \\ 
	 	 \hline 
	 \end{tabular} 
\end{table*} 

%% file: figures/statistics_as_latex-3.txt
april-19-2024-A2
\begin{table*}
	 \centering
	 \caption{SOM Clustering for April 19 with Antenna A2.}
	 \label{tab:[]}
	 \begin{tabular}{cclllll}
	 	 \hline
	 	 Cluster \# & \# Pulses & Peak Loc & Peak Height & Peak Width & Peak Skew & MSE \\ 
	 	 \hline 
	 	 0 & 147363 & $100.64 \pm 2.32$ & $8.56 \pm 6.14$ & $5.88 \pm 2.17$ & $4.46 \pm 0.93$ & $0.00004 \pm 0.00006$  \\ 
	 	 1 & 411 & $96.14 \pm 0.56$ & $69.71 \pm 31.92$ & $3.19 \pm 0.34$ & $6.99 \pm 0.54$ & $0.02110 \pm 0.09954$  \\ 
	 	 2 & 1790 & $97.64 \pm 0.73$ & $32.69 \pm 10.67$ & $3.57 \pm 0.52$ & $5.97 \pm 0.65$ & $0.00363 \pm 0.00544$  \\ 
	 	 3 & 18649 & $98.77 \pm 1.36$ & $15.21 \pm 3.76$ & $5.62 \pm 1.82$ & $4.98 \pm 0.71$ & $0.00034 \pm 0.00051$  \\ 
	 	 4 & 54926 & $99.99 \pm 1.80$ & $8.96 \pm 2.35$ & $5.35 \pm 1.87$ & $4.51 \pm 0.76$ & $0.00011 \pm 0.00016$  \\ 
	 	 5 & 69839 & $101.74 \pm 2.31$ & $5.56 \pm 2.06$ & $5.19 \pm 1.68$ & $4.24 \pm 0.99$ & $0.00008 \pm 0.00012$  \\ 
	 	 6 & 1748 & $101.03 \pm 2.22$ & $5.45 \pm 3.01$ & $4.59 \pm 1.36$ & $4.09 \pm 1.06$ & $0.00375 \pm 0.00582$  \\ 
	 	 \hline 
	 \end{tabular} 
\end{table*} 

%% file: figures/statistics_as_latex-4.txt
april-20-2024-A2
\begin{table*}
	 \centering
	 \caption{SOM Clustering for April 20 with Antenna A2.}
	 \label{tab:[]}
	 \begin{tabular}{cclllll}
	 	 \hline
	 	 Cluster \# & \# Pulses & Peak Loc & Peak Height & Peak Width & Peak Skew & MSE \\ 
	 	 \hline 
	 	 0 & 147400 & $100.32 \pm 2.71$ & $9.17 \pm 6.29$ & $5.91 \pm 2.17$ & $4.59 \pm 0.89$ & $0.00004 \pm 0.00007$  \\ 
	 	 1 & 169 & $95.41 \pm 0.64$ & $92.95 \pm 32.24$ & $3.42 \pm 0.11$ & $6.83 \pm 0.18$ & $0.05770 \pm 0.22798$  \\ 
	 	 2 & 1465 & $96.68 \pm 0.70$ & $39.78 \pm 12.67$ & $3.79 \pm 0.46$ & $6.09 \pm 0.58$ & $0.00494 \pm 0.00783$  \\ 
	 	 3 & 103 & $97.09 \pm 3.27$ & $24.55 \pm 9.75$ & $4.63 \pm 2.26$ & $4.85 \pm 1.38$ & $0.11536 \pm 0.22739$  \\ 
	 	 4 & 16570 & $97.85 \pm 1.18$ & $16.69 \pm 4.55$ & $4.87 \pm 1.30$ & $5.02 \pm 0.72$ & $0.00041 \pm 0.00061$  \\ 
	 	 5 & 64490 & $99.80 \pm 1.74$ & $9.85 \pm 2.77$ & $5.27 \pm 1.70$ & $4.58 \pm 0.79$ & $0.00010 \pm 0.00015$  \\ 
	 	 6 & 64603 & $101.56 \pm 3.10$ & $5.63 \pm 1.82$ & $5.00 \pm 1.53$ & $4.46 \pm 0.97$ & $0.00010 \pm 0.00015$  \\ 
	 	 \hline 
	 \end{tabular} 
\end{table*} 

%% file: figures/statistics_as_latex-5.txt
april-28-2024-A2
\begin{table*}
	 \centering
	 \caption{SOM Clustering for April 28 with Antenna A2.}
	 \label{tab:[]}
	 \begin{tabular}{cclllll}
	 	 \hline
	 	 Cluster \# & \# Pulses & Peak Loc & Peak Height & Peak Width & Peak Skew & MSE \\ 
	 	 \hline 
	 	 0 & 144793 & $100.77 \pm 2.92$ & $8.19 \pm 5.51$ & $5.68 \pm 2.19$ & $4.58 \pm 1.01$ & $0.00005 \pm 0.00007$  \\ 
	 	 1 & 202 & $95.81 \pm 0.61$ & $85.93 \pm 22.61$ & $3.27 \pm 0.44$ & $6.96 \pm 0.53$ & $0.05445 \pm 0.21490$  \\ 
	 	 2 & 1416 & $96.65 \pm 0.75$ & $34.37 \pm 10.69$ & $3.43 \pm 0.61$ & $6.23 \pm 0.80$ & $0.00511 \pm 0.00773$  \\ 
	 	 3 & 32808 & $98.33 \pm 1.49$ & $11.60 \pm 4.14$ & $6.28 \pm 2.34$ & $4.75 \pm 0.84$ & $0.00021 \pm 0.00031$  \\ 
	 	 4 & 245 & $101.67 \pm 3.94$ & $10.22 \pm 3.91$ & $6.75 \pm 3.00$ & $2.56 \pm 0.85$ & $0.03333 \pm 0.05936$  \\ 
	 	 5 & 47830 & $101.31 \pm 1.91$ & $8.60 \pm 2.35$ & $5.32 \pm 1.88$ & $4.49 \pm 0.84$ & $0.00014 \pm 0.00021$  \\ 
	 	 6 & 62292 & $101.75 \pm 3.32$ & $5.22 \pm 1.64$ & $3.69 \pm 0.76$ & $4.53 \pm 1.16$ & $0.00011 \pm 0.00016$  \\ 
	 	 \hline 
	 \end{tabular} 
\end{table*} 

%% file: figures/statistics_as_latex-6.txt
may-01-2024-A2
\begin{table*}
	 \centering
	 \caption{SOM Clustering for May 01 with Antenna A2.}
	 \label{tab:[]}
	 \begin{tabular}{cclllll}
	 	 \hline
	 	 Cluster \# & \# Pulses & Peak Loc & Peak Height & Peak Width & Peak Skew & MSE \\ 
	 	 \hline 
	 	 0 & 144791 & $103.73 \pm 2.72$ & $9.59 \pm 6.34$ & $4.73 \pm 1.46$ & $4.65 \pm 0.94$ & $0.00005 \pm 0.00007$  \\ 
	 	 1 & 739 & $99.57 \pm 0.77$ & $62.91 \pm 26.85$ & $3.38 \pm 0.52$ & $6.71 \pm 0.67$ & $0.01006 \pm 0.01723$  \\ 
	 	 2 & 4366 & $99.92 \pm 1.05$ & $21.01 \pm 6.91$ & $4.25 \pm 1.20$ & $5.24 \pm 0.84$ & $0.00153 \pm 0.00225$  \\ 
	 	 3 & 10619 & $101.86 \pm 0.90$ & $18.07 \pm 4.84$ & $4.67 \pm 1.24$ & $5.14 \pm 0.74$ & $0.00063 \pm 0.00093$  \\ 
	 	 4 & 36765 & $103.42 \pm 1.27$ & $10.82 \pm 2.45$ & $4.29 \pm 1.07$ & $4.89 \pm 0.73$ & $0.00018 \pm 0.00026$  \\ 
	 	 5 & 23343 & $102.35 \pm 2.70$ & $9.63 \pm 2.35$ & $6.87 \pm 2.82$ & $4.18 \pm 0.70$ & $0.00028 \pm 0.00042$  \\ 
	 	 6 & 68959 & $104.95 \pm 2.86$ & $6.33 \pm 1.93$ & $4.41 \pm 1.59$ & $4.55 \pm 1.03$ & $0.00010 \pm 0.00014$  \\ 
	 	 \hline 
	 \end{tabular} 
\end{table*} 

%% file: figures/statistics_as_latex-7.txt
may-02-2024-A2
\begin{table*}
	 \centering
	 \caption{SOM Clustering for May 02 with Antenna A2.}
	 \label{tab:[]}
	 \begin{tabular}{cclllll}
	 	 \hline
	 	 Cluster \# & \# Pulses & Peak Loc & Peak Height & Peak Width & Peak Skew & MSE \\ 
	 	 \hline 
	 	 0 & 147394 & $100.84 \pm 2.83$ & $9.65 \pm 6.93$ & $5.53 \pm 2.13$ & $4.66 \pm 0.80$ & $0.00004 \pm 0.00006$  \\ 
	 	 1 & 280 & $95.95 \pm 0.67$ & $100.70 \pm 26.99$ & $3.15 \pm 0.33$ & $7.15 \pm 0.50$ & $0.03478 \pm 0.15753$  \\ 
	 	 2 & 1219 & $96.59 \pm 0.75$ & $43.15 \pm 12.08$ & $3.41 \pm 0.59$ & $6.36 \pm 0.77$ & $0.00537 \pm 0.00795$  \\ 
	 	 3 & 9171 & $98.15 \pm 1.12$ & $20.58 \pm 5.37$ & $4.12 \pm 1.01$ & $5.24 \pm 0.76$ & $0.00069 \pm 0.00101$  \\ 
	 	 4 & 23255 & $98.48 \pm 1.79$ & $11.71 \pm 2.97$ & $6.07 \pm 2.35$ & $4.65 \pm 0.73$ & $0.00027 \pm 0.00040$  \\ 
	 	 5 & 54662 & $101.23 \pm 1.57$ & $9.61 \pm 2.63$ & $5.89 \pm 2.16$ & $4.67 \pm 0.73$ & $0.00011 \pm 0.00017$  \\ 
	 	 6 & 58807 & $101.96 \pm 3.34$ & $6.05 \pm 1.85$ & $4.17 \pm 1.69$ & $4.52 \pm 0.81$ & $0.00011 \pm 0.00016$  \\ 
	 	 \hline 
	 \end{tabular} 
\end{table*} 

%% file: figures/statistics_as_latex-8.txt
may-03-2024-A2
\begin{table*}
	 \centering
	 \caption{SOM Clustering for May 03 with Antenna A2.}
	 \label{tab:[]}
	 \begin{tabular}{cclllll}
	 	 \hline
	 	 Cluster \# & \# Pulses & Peak Loc & Peak Height & Peak Width & Peak Skew & MSE \\ 
	 	 \hline 
	 	 0 & 147393 & $100.82 \pm 2.65$ & $10.39 \pm 6.97$ & $5.99 \pm 2.30$ & $4.63 \pm 0.85$ & $0.00004 \pm 0.00006$  \\ 
	 	 1 & 134 & $95.84 \pm 0.57$ & $113.98 \pm 24.09$ & $3.24 \pm 0.25$ & $7.11 \pm 0.54$ & $0.07795 \pm 0.32802$  \\ 
	 	 2 & 792 & $96.47 \pm 0.66$ & $55.08 \pm 14.63$ & $3.22 \pm 0.42$ & $6.63 \pm 0.71$ & $0.00871 \pm 0.01335$  \\ 
	 	 3 & 8875 & $97.48 \pm 1.10$ & $22.15 \pm 7.02$ & $5.03 \pm 1.63$ & $5.21 \pm 0.80$ & $0.00073 \pm 0.00108$  \\ 
	 	 4 & 29016 & $99.38 \pm 1.19$ & $14.40 \pm 3.50$ & $5.39 \pm 1.78$ & $4.94 \pm 0.76$ & $0.00022 \pm 0.00033$  \\ 
	 	 5 & 60630 & $101.02 \pm 2.17$ & $9.35 \pm 2.11$ & $6.11 \pm 2.23$ & $4.47 \pm 0.70$ & $0.00010 \pm 0.00016$  \\ 
	 	 6 & 47946 & $102.14 \pm 3.02$ & $6.07 \pm 1.88$ & $4.76 \pm 1.84$ & $4.50 \pm 0.93$ & $0.00013 \pm 0.00020$  \\ 
	 	 \hline 
	 \end{tabular} 
\end{table*} 

%% file: figures/statistics_as_latex-9.txt
may-04-2024-A2
\begin{table*}
	 \centering
	 \caption{SOM Clustering for May 04 with Antenna A2.}
	 \label{tab:[]}
	 \begin{tabular}{cclllll}
	 	 \hline
	 	 Cluster \# & \# Pulses & Peak Loc & Peak Height & Peak Width & Peak Skew & MSE \\ 
	 	 \hline 
	 	 0 & 142092 & $100.85 \pm 2.68$ & $11.08 \pm 7.44$ & $5.90 \pm 2.27$ & $4.66 \pm 0.76$ & $0.00005 \pm 0.00007$  \\ 
	 	 1 & 238 & $96.02 \pm 0.65$ & $105.71 \pm 29.24$ & $3.18 \pm 0.32$ & $7.08 \pm 0.54$ & $0.04030 \pm 0.09199$  \\ 
	 	 2 & 1941 & $97.03 \pm 0.74$ & $41.17 \pm 12.85$ & $3.76 \pm 0.79$ & $6.04 \pm 0.74$ & $0.00370 \pm 0.00557$  \\ 
	 	 3 & 8548 & $98.74 \pm 1.01$ & $22.46 \pm 4.90$ & $3.55 \pm 0.50$ & $5.27 \pm 0.69$ & $0.00082 \pm 0.00122$  \\ 
	 	 4 & 23413 & $98.41 \pm 1.70$ & $13.62 \pm 3.54$ & $5.60 \pm 2.06$ & $4.63 \pm 0.68$ & $0.00030 \pm 0.00044$  \\ 
	 	 5 & 52190 & $100.95 \pm 1.59$ & $11.09 \pm 2.95$ & $5.55 \pm 1.85$ & $4.64 \pm 0.67$ & $0.00013 \pm 0.00020$  \\ 
	 	 6 & 55762 & $102.26 \pm 2.96$ & $6.81 \pm 2.04$ & $5.18 \pm 2.18$ & $4.54 \pm 0.77$ & $0.00012 \pm 0.00018$  \\ 
	 	 \hline 
	 \end{tabular} 
\end{table*} 

%% file: paper.bbl
\begin{thebibliography}{57}
\expandafter\ifx\csname natexlab\endcsname\relax\def\natexlab#1{#1}\fi

\bibitem[{{Alpar} {et~al.}(1993){Alpar}, {Chau}, {Cheng}, \&
  {Pines}}]{1993ApJ...409..345A}
{Alpar}, M.~A., {Chau}, H.~F., {Cheng}, K.~S., \& {Pines}, D. 1993, \apj, 409,
  345

\bibitem[{{Andersson} {et~al.}(2012){Andersson}, {Glampedakis}, {Ho}, \&
  {Espinoza}}]{2012PhRvL.109x1103A}
{Andersson}, N., {Glampedakis}, K., {Ho}, W.~C.~G., \& {Espinoza}, C.~M. 2012,
  \prl, 109, 241103

\bibitem[{{Antonopoulou} {et~al.}(2022){Antonopoulou}, {Haskell}, \&
  {Espinoza}}]{2022RPPh...85l6901A}
{Antonopoulou}, D., {Haskell}, B., \& {Espinoza}, C.~M. 2022, Reports on
  Progress in Physics, 85, 126901

\bibitem[{{Araujo Furlan} {et~al.}(2023){Araujo Furlan}, {Gancio}, {Galante},
  \& {Romero}}]{2023BAAA...64..304A}
{Araujo Furlan}, S.~B., {Gancio}, G., {Galante}, C.~A., \& {Romero}, G.~E.
  2023, Boletin de la Asociacion Argentina de Astronomia La Plata Argentina,
  64, 304

\bibitem[{{Ashton} {et~al.}(2019){Ashton}, {Lasky}, {Graber}, \&
  {Palfreyman}}]{2019NatAs...3.1143A}
{Ashton}, G., {Lasky}, P.~D., {Graber}, V., \& {Palfreyman}, J. 2019, Nature
  Astronomy, 3, 1143

\bibitem[{{Basu} {et~al.}(2022){Basu}, {Shaw}, {Antonopoulou}, {Keith}, {Lyne},
  {Mickaliger}, {Stappers}, {Weltevrede}, \& {Jordan}}]{2022MNRAS.510.4049B}
{Basu}, A., {Shaw}, B., {Antonopoulou}, D., {et~al.} 2022, \mnras, 510, 4049

\bibitem[{{Bransgrove} {et~al.}(2020){Bransgrove}, {Beloborodov}, \&
  {Levin}}]{2020ApJ...897..173B}
{Bransgrove}, A., {Beloborodov}, A.~M., \& {Levin}, Y. 2020, \apj, 897, 173

\bibitem[{{Cairns} {et~al.}(2001){Cairns}, {Johnston}, \& {Das}}]{Cairns+2001}
{Cairns}, I.~H., {Johnston}, S., \& {Das}, P. 2001, \apjl, 563, L65

\bibitem[{{Chamel}(2013)}]{2013PhRvL.110a1101C}
{Chamel}, N. 2013, \prl, 110, 011101

\bibitem[{{Dodson} {et~al.}(2002){Dodson}, {McCulloch}, \&
  {Lewis}}]{2002ApJ...564L..85D}
{Dodson}, R.~G., {McCulloch}, P.~M., \& {Lewis}, D.~R. 2002, \apjl, 564, L85

\bibitem[{{Espinoza} {et~al.}(2011){Espinoza}, {Lyne}, {Stappers}, \&
  {Kramer}}]{Espinoza:2011pq}
{Espinoza}, C.~M., {Lyne}, A.~G., {Stappers}, B.~W., \& {Kramer}, M. 2011,
  \mnras, 414, 1679

\bibitem[{Flanagan(1990)}]{Flanagan1990RapidRO}
Flanagan, C.~S. 1990, Nature, 345, 416

\bibitem[{{Gancio} {et~al.}(2020){Gancio}, {Lousto}, {Combi}, {del Palacio},
  {L{\'o}pez Armengol}, {Combi}, {Garc{\'\i}a}, {Kornecki}, {M{\"u}ller},
  {Guti{\'e}rrez}, {Hauscarriaga}, \& {Mancuso}}]{Gancio2020}
{Gancio}, G., {Lousto}, C.~O., {Combi}, L., {et~al.} 2020, \aap, 633, A84

\bibitem[{{Gancio} {et~al.}(2024){Gancio}, {Romero}, {Astudillo}, {Saavedra},
  \& {Combi}}]{2024RMxAC..56..131G}
{Gancio}, G., {Romero}, G.~E., {Astudillo}, J., {Saavedra}, E.~A., \& {Combi},
  J.~A. 2024, in Revista Mexicana de Astronomia y Astrofisica Conference
  Series, Vol.~56, Revista Mexicana de Astronomia y Astrofisica Conference
  Series, 131--133

\bibitem[{{Graber} {et~al.}(2018){Graber}, {Cumming}, \&
  {Andersson}}]{2018ApJ...865...23G}
{Graber}, V., {Cumming}, A., \& {Andersson}, N. 2018, \apj, 865, 23

\bibitem[{{G{\"u}gercino{\u{g}}lu}(2017)}]{2017JPhCS.932a2037G}
{G{\"u}gercino{\u{g}}lu}, E. 2017, in Journal of Physics Conference Series,
  Vol. 932, Journal of Physics Conference Series (IOP), 012037

\bibitem[{{G{\"u}gercino{\u{g}}lu} \& {Alpar}(2020)}]{2020MNRAS.496.2506G}
{G{\"u}gercino{\u{g}}lu}, E. \& {Alpar}, M.~A. 2020, \mnras, 496, 2506

\bibitem[{{Haskell} \& {Melatos}(2015)}]{2015IJMPD..2430008H}
{Haskell}, B. \& {Melatos}, A. 2015, International Journal of Modern Physics D,
  24, 1530008

\bibitem[{{Ho} {et~al.}(2015){Ho}, {Espinoza}, {Antonopoulou}, \&
  {Andersson}}]{2015SciA....1E0578H}
{Ho}, W.~C.~G., {Espinoza}, C.~M., {Antonopoulou}, D., \& {Andersson}, N. 2015,
  Science Advances, 1, e1500578

\bibitem[{{Hobbs} {et~al.}(2012){Hobbs}, {Coles}, {Manchester}, {Keith},
  {Shannon}, {Chen}, {Bailes}, {Bhat}, {Burke-Spolaor}, {Champion},
  {Chaudhary}, {Hotan}, {Khoo}, {Kocz}, {Levin}, {Oslowski}, {Preisig}, {Ravi},
  {Reynolds}, {Sarkissian}, {van Straten}, {Verbiest}, {Yardley}, \&
  {You}}]{2012MNRAS.427.2780H}
{Hobbs}, G., {Coles}, W., {Manchester}, R.~N., {et~al.} 2012, \mnras, 427, 2780

\bibitem[{{Hotan} {et~al.}(2004){Hotan}, {van Straten}, \&
  {Manchester}}]{2004PASA...21..302H}
{Hotan}, A.~W., {van Straten}, W., \& {Manchester}, R.~N. 2004, \pasa, 21, 302

\bibitem[{Johnston {et~al.}(2001)Johnston, van Straten, Kramer, \&
  Bailes}]{Johnston+2001}
Johnston, S., van Straten, W., Kramer, M., \& Bailes, M. 2001, The
  Astrophysical Journal, 549, L101

\bibitem[{Khomenko \& Haskell(2018)}]{khomenko_haskell_2018}
Khomenko, V. \& Haskell, B. 2018, Publications of the Astronomical Society of
  Australia, 35, e020

\bibitem[{Kingma \& Welling(2014)}]{kingma2014autoencoding}
Kingma, D.~P. \& Welling, M. 2014, Auto-Encoding Variational Bayes

\bibitem[{Kohonen(1988)}]{teuvo1988som}
Kohonen, T. 1988, Self-Organized Formation of Topologically Correct Feature
  Maps (Cambridge, MA, USA: MIT Press), 509–521

\bibitem[{{Lentati} {et~al.}(2014){Lentati}, {Alexander}, {Hobson}, {Feroz},
  {van Haasteren}, {Lee}, \& {Shannon}}]{2014MNRAS.437.3004L}
{Lentati}, L., {Alexander}, P., {Hobson}, M.~P., {et~al.} 2014, \mnras, 437,
  3004

\bibitem[{{Link} {et~al.}(1999){Link}, {Epstein}, \&
  {Lattimer}}]{1999PhRvL..83.3362L}
{Link}, B., {Epstein}, R.~I., \& {Lattimer}, J.~M. 1999, \prl, 83, 3362

\bibitem[{{Lopez Armengol} {et~al.}(2019{\natexlab{a}}){Lopez Armengol},
  {Lousto}, {del Palacio}, {Garcia}, {Combi}, {Combi}, {Gancio}, {Mueller}, \&
  {Kornecki}}]{atel_vela}
{Lopez Armengol}, F.~G., {Lousto}, C.~O., {del Palacio}, S., {et~al.}
  2019{\natexlab{a}}, The Astronomer's Telegram, 12482, 1

\bibitem[{{Lopez Armengol} {et~al.}(2019{\natexlab{b}}){Lopez Armengol},
  {Lousto}, {del Palacio}, {Garcia}, {Combi}, {Combi}, {Gancio}, {Mueller}, \&
  {Kornecki}}]{2019ATel12482....1L}
{Lopez Armengol}, F.~G., {Lousto}, C.~O., {del Palacio}, S., {et~al.}
  2019{\natexlab{b}}, The Astronomer's Telegram, 12482, 1

\bibitem[{Lousto {et~al.}(2021)Lousto, Missel, Prajapati, Fiscella, Armengol,
  Gyawali, Wang, Cahill, Combi, del Palacio, Combi, Gancio, García,
  Gutiérrez, \& Hauscarriaga}]{Lousto:2021dia}
Lousto, C.~O., Missel, R., Prajapati, H., {et~al.} 2021, Monthly Notices of the
  Royal Astronomical Society
  [\eprint{https://academic.oup.com/mnras/advance-article-pdf/doi/10.1093/mnras/stab3287/41243307/stab3287.pdf}],
  stab3287

\bibitem[{{Lyne}(1992)}]{1992RSPTA.341...29L}
{Lyne}, A.~G. 1992, Philosophical Transactions of the Royal Society of London
  Series A, 341, 29

\bibitem[{{Maan} {et~al.}(2021){Maan}, {van Leeuwen}, \& {Vohl}}]{Maan2020}
{Maan}, Y., {van Leeuwen}, J., \& {Vohl}, D. 2021, \aap, 650, A80

\bibitem[{Mahida {et~al.}(2023)Mahida, Palfreyman, Calves, \&
  Sett}]{Mahida:2023maa}
Mahida, A.~D., Palfreyman, J.~L., Calves, G.~M., \& Sett, S. 2023, Mon. Not.
  Roy. Astron. Soc., 524, 759

\bibitem[{{Manchester}(2018)}]{Manchester:2018jhy}
{Manchester}, R.~N. 2018, in IAU Symposium, Vol. 337, Pulsar Astrophysics the
  Next Fifty Years, ed. P.~{Weltevrede}, B.~B.~P. {Perera}, L.~L. {Preston}, \&
  S.~{Sanidas}, 197--202

\bibitem[{{Manchester} {et~al.}(2005){Manchester}, {Hobbs}, {Teoh}, \&
  {Hobbs}}]{ManchesterATNF2005}
{Manchester}, R.~N., {Hobbs}, G.~B., {Teoh}, A., \& {Hobbs}, M. 2005, \aj, 129,
  1993

\bibitem[{Mcculloch {et~al.}(1987)Mcculloch, Klekociuk, Hamilton, \&
  Royle}]{glitch-timing}
Mcculloch, P., Klekociuk, A., Hamilton, P., \& Royle, G. 1987, Australian
  Journal of Physics, 40, 725

\bibitem[{{Montoli} {et~al.}(2020){Montoli}, {Antonelli}, \&
  {Pizzochero}}]{2020MNRAS.492.4837M}
{Montoli}, A., {Antonelli}, M., \& {Pizzochero}, P.~M. 2020, \mnras, 492, 4837

\bibitem[{{Palfreyman}(2024)}]{2024ATel16615....1P}
{Palfreyman}, J. 2024, The Astronomer's Telegram, 16615, 1

\bibitem[{{Palfreyman} {et~al.}(2018){Palfreyman}, {Dickey}, {Hotan},
  {Ellingsen}, \& {van Straten}}]{2018Natur.556..219P}
{Palfreyman}, J., {Dickey}, J.~M., {Hotan}, A., {Ellingsen}, S., \& {van
  Straten}, W. 2018, \nat, 556, 219

\bibitem[{{Palfreyman} {et~al.}(2016){Palfreyman}, {Dickey}, {Ellingsen},
  {Jones}, \& {Hotan}}]{Palfreyman+2016}
{Palfreyman}, J.~L., {Dickey}, J.~M., {Ellingsen}, S.~P., {Jones}, I.~R., \&
  {Hotan}, A.~W. 2016, \apj, 820, 64

\bibitem[{{Press} {et~al.}(1992){Press}, {Teukolsky}, {Vetterling}, \&
  {Flannery}}]{Press1992}
{Press}, W.~H., {Teukolsky}, S.~A., {Vetterling}, W.~T., \& {Flannery}, B.~P.
  1992, {Numerical recipes in C. The art of scientific computing} ({IOP}
  Publishing)

\bibitem[{{Radhakrishnan} \& {Manchester}(1969)}]{1969Natur.222..228R}
{Radhakrishnan}, V. \& {Manchester}, R.~N. 1969, \nat, 222, 228

\bibitem[{{Ransom}(2011)}]{2011ascl.soft07017R}
{Ransom}, S. 2011, {PRESTO: PulsaR Exploration and Search TOolkit}

\bibitem[{{Ransom}(2018)}]{PRESTO}
{Ransom}, S. 2018, PRESTO - Pulsar Exploration and Search Toolkit

\bibitem[{{Reichley} \& {Downs}(1969)}]{1969Natur.222..229R}
{Reichley}, P.~E. \& {Downs}, G.~S. 1969, \nat, 222, 229

\bibitem[{{Sosa-Fiscella} {et~al.}(2021){Sosa-Fiscella}, {Zubieta}, {del
  Palacio}, {Garcia}, {Lopez-Armengol}, {Combi}, {Lousto}, {Gancio}, {Combi},
  {Gutierrez}, {Bunzel}, {Hauscarriaga}, \& {PuMA
  Collaboration}}]{2021ATel14806....1S}
{Sosa-Fiscella}, V., {Zubieta}, E., {del Palacio}, S., {et~al.} 2021, The
  Astronomer's Telegram, 14806, 1

\bibitem[{{Taylor}(1992)}]{1992PTRSL.341..117T}
{Taylor}, J.~H. 1992, Philosophical Transactions of the Royal Society of
  London, 341, 117

\bibitem[{{Yu} {et~al.}(2013{\natexlab{a}}){Yu}, {Manchester}, {Hobbs},
  {Johnston}, {Kaspi}, {Keith}, {Lyne}, {Qiao}, {Ravi}, {Sarkissian},
  {Shannon}, \& {Xu}}]{Yu2013}
{Yu}, M., {Manchester}, R.~N., {Hobbs}, G., {et~al.} 2013{\natexlab{a}},
  \mnras, 429, 688

\bibitem[{{Yu} {et~al.}(2013{\natexlab{b}}){Yu}, {Manchester}, {Hobbs},
  {Johnston}, {Kaspi}, {Keith}, {Lyne}, {Qiao}, {Ravi}, {Sarkissian},
  {Shannon}, \& {Xu}}]{2013MNRAS.429..688Y}
{Yu}, M., {Manchester}, R.~N., {Hobbs}, G., {et~al.} 2013{\natexlab{b}},
  \mnras, 429, 688

\bibitem[{{Zubieta} {et~al.}(2024{\natexlab{a}}){Zubieta}, {Araujo Furlan},
  {del Palacio}, {Garcia}, {Gancio}, {Lousto}, \&
  {Combi}}]{2024ATel16580....1Z}
{Zubieta}, E., {Araujo Furlan}, S.~B., {del Palacio}, S., {et~al.}
  2024{\natexlab{a}}, The Astronomer's Telegram, 16580, 1

\bibitem[{{Zubieta} {et~al.}(2024{\natexlab{b}}){Zubieta}, {del Palacio},
  {Garc{\'\i}a}, {Araujo Furlan}, {Gancio}, {Lousto}, \&
  {Combi}}]{2024RMxAC..56..161Z}
{Zubieta}, E., {del Palacio}, S., {Garc{\'\i}a}, F., {et~al.}
  2024{\natexlab{b}}, in Revista Mexicana de Astronomia y Astrofisica
  Conference Series, Vol.~56, Revista Mexicana de Astronomia y Astrofisica
  Conference Series, 161--165

\bibitem[{{Zubieta} {et~al.}(2022{\natexlab{a}}){Zubieta}, {Del Palacio},
  {Garcia}, {Gancio}, {Lousto}, {Combi}, {Combi}, {Gutierrez},
  {Lopez-Armengol}, {Simaz Bunzel}, \& {Sosa-Fiscella}}]{2022ATel15638....1Z}
{Zubieta}, E., {Del Palacio}, S., {Garcia}, F., {et~al.} 2022{\natexlab{a}},
  The Astronomer's Telegram, 15638, 1

\bibitem[{{Zubieta} {et~al.}(2022{\natexlab{b}}){Zubieta}, {Furlan}, {Palacio},
  {Garcia}, {Gancio}, {Lousto}, {Combi}, \& {Combi}}]{2022ATel15838....1Z}
{Zubieta}, E., {Furlan}, S.~B.~A., {Palacio}, S.~d., {et~al.}
  2022{\natexlab{b}}, The Astronomer's Telegram, 15838, 1

\bibitem[{{Zubieta} {et~al.}(2024{\natexlab{c}}){Zubieta}, {Furlan}, {Palacio},
  {Garcia}, {Gancio}, {Lousto}, {Combi}, \& {PuMA
  Collaboration}}]{2024ATel16608....1Z}
{Zubieta}, E., {Furlan}, S.~B.~A., {Palacio}, S.~d., {et~al.}
  2024{\natexlab{c}}, The Astronomer's Telegram, 16608, 1

\bibitem[{{Zubieta} {et~al.}(2024{\natexlab{d}}){Zubieta}, {Garc{\'\i}a}, {del
  Palacio}, {Araujo Furlan}, {Gancio}, {Lousto}, {Combi}, \&
  {Espinoza}}]{2024A&A...689A.191Z}
{Zubieta}, E., {Garc{\'\i}a}, F., {del Palacio}, S., {et~al.}
  2024{\natexlab{d}}, \aap, 689, A191

\bibitem[{{Zubieta} {et~al.}(2024{\natexlab{e}}){Zubieta}, {Garc{\'\i}a}, {del
  Palacio}, {Espinoza}, {Araujo Furlan}, {Gancio}, {Lousto}, {Combi}, \&
  {G{\"u}gercino{\u{g}}lu}}]{2024arXiv241217766Z}
{Zubieta}, E., {Garc{\'\i}a}, F., {del Palacio}, S., {et~al.}
  2024{\natexlab{e}}, arXiv e-prints, arXiv:2412.17766

\bibitem[{Zubieta {et~al.}(2023)}]{Zubieta:2022umm}
Zubieta, E. {et~al.} 2023, Mon. Not. Roy. Astron. Soc., 521, 4504

\end{thebibliography}
